\documentclass[preprint,3p]{elsarticle}
\usepackage[toc,page,title,titletoc,header]{appendix}
\usepackage{stmaryrd}
\SetSymbolFont{stmry}{bold}{U}{stmry}{m}{n}
\usepackage{amsmath}
\usepackage{amssymb}
\usepackage{amsthm}
\usepackage{tabularx}
\usepackage{subfigure}
\usepackage{epsfig}
\usepackage{indentfirst}
\usepackage{cases}
\biboptions{sort&compress}
\usepackage{graphicx}
\usepackage{color}
\usepackage{float}
\usepackage{multirow}
\usepackage{subfigure}

\renewcommand{\vec}[1]{\mathbf{#1}}

\def\subFV{\scriptscriptstyle{FV}}
\def\subFS{\scriptscriptstyle{FS}}
\def\subVS{\scriptscriptstyle{VS}}

\newtheorem{thm}{Theorem}[section]
\newtheorem{prop}{Proposition}[section]

\newtheorem{lem}{Lemma}[section]

\numberwithin{equation}{section}

\begin{document}

\begin{frontmatter}

\title{Sharp-interface approach for simulating solid-state dewetting in two dimensions: a Cahn-Hoffman $\boldsymbol{\xi}$-vector formulation}

\author[1]{Wei Jiang}
\address[1]{School of Mathematics and Statistics,
Wuhan University, Wuhan, 430072, China}
\ead{jiangwei1007@whu.edu.cn}

\author[2]{Quan Zhao\corref{3}}
\address[2]{Department of Mathematics, National University of
Singapore, 119076, Singapore}
\ead{quanzhao90@u.nus.edu}
\cortext[3]{Corresponding author.}


\begin{abstract}
By using a Cahn-Hoffman $\boldsymbol{\xi}$-vector formulation, we propose a sharp-interface approach for solving solid-state dewetting problems in two dimensions. First, based on the thermodynamic variation and smooth vector-field perturbation method, we rigorously derive a sharp-interface model with weakly anisotropic surface energies, and this model describes the interface evolution which occurs through surface diffusion flow and contact line migration. Second, a parametric finite element method in terms of the $\boldsymbol{\xi}$-vector formulation is proposed for numerically solving the sharp-interface model. By performing numerical simulations, we examine several specific evolution processes for solid-state dewetting of thin films, e.g., the evolution of small islands, pinch-off of large islands and power-law retraction dynamics of semi-infinite step films, and these simulation results are highly consistent with experimental observations. Finally, we also include the strong surface energy anisotropy into the sharp-interface model and design its corresponding numerical scheme via the $\boldsymbol{\xi}$-vector formulation.

\end{abstract}



\begin{keyword}
Solid-state dewetting, surface energy anisotropy, surface diffusion, moving contact lines, Cahn-Hoffman $\boldsymbol{\xi}$-vector.
\end{keyword}

\end{frontmatter}

\section{Introduction}

Solid-state dewetting is a ubiquitous phenomenon in physics and materials science. It occurs in the solid-solid-vapor system and describes the agglomeration of solid thin films on a substrate (e.g., see the recent review papers \cite{Thompson12,Leroy16}). In general, thin solid films rested on substrates are rarely stable
and will undergo particle formation (often called as ``dewetting'' or agglomeration) due to surface tension/capillarity effects, and it could exhibit complex features during the evolution, such as the faceting~\cite{Jiran90,Jiran92,Ye10a}, edge retraction~\cite{Wong00,Dornel06,Kim13}, pinch-off events~\cite{Jiang12,Kim15}, fingering instabilities~\cite{Kan05,Ye10b,Ye11a,Ye11b} and so on. Recently, solid-state dewetting has demonstrated its wide applications in many modern technologies where it can be either detrimental or advantageous. For example, it can destroy micro-/nanodevice performance through surface instabilities in well-prepared patterned structure, and  on the other hand, it can be positively used to produce well-controlled patterns of an array of micro-/nanoscale particles, which can be used for sensors~\cite{Mizsei93}, for optical and magnetic devices~\cite{Armelao06}, and as catalysts for the growth of carbon and semiconductor nanowire growth~\cite{Schmidt09}. Therefore, it is very necessary to develop suitable theoretical models for investigating the kinetic evolution process and understanding its intrinsic physical laws during solid-state dewetting.

Recently, solid-state dewetting of thin films has attracted considerable interests, and it has been studied experimentally (e.g.,~\cite{Jiran90,Jiran92,Ye10a,Ye10b,Ye11a,Amram12,Rabkin14,Herz216,Naffouti16,Naffouti17,Kovalenko17}) and theoretically (e.g.,~\cite{Jiang12,Srolovitz86a,Srolovitz86,Wang15,Jiang16,Bao17,Bao17b,Dornel06,Wong00,Kim13,Kim15,Kan05,Zucker16}) by many research groups. Different from traditional wetting/dewetting phenomena (i.e., ``liquid-state'' dewetting), mass transport during solid-state dewetting is usually dominated by surface diffusion rather than fluid dynamics. It can be thought of as a kind of interface-tracking problems where the morphology evolution is governed by surface diffusion flow and contact line migration~\cite{Jiang12,Wang15,Bao17}. In addition, unlike liquid or amorphous solids, the surface energy (density) of thin film materials often exhibits strong dependence on its crystallographic orientations. This property is usually called as ``surface energy anisotropy'' in literatures, and it could greatly influence the kinetics and morphology evolution during solid-state dewetting, and result in many interesting phenomena such as the faceting~\cite{Thompson12,Ye10a,Ye11a}, which have been usually observed in real physical experiments. To model anisotropic thin film solid materials, the surface energy density (labeled as $\gamma$), is often assumed to be a function of the unit normal direction $\vec n$ of the interface curve/surface, and it can be expressed as a continuous positive function defined on the unit circle $S^1$ in two dimensions:
\begin{equation}
\gamma(\vec n):\quad S^1 \rightarrow  \mathbb{R}^+,
\end{equation}
where $\vec n$ is the unit outer normal vector of the crystalline interface.
\begin{figure}[!htp]
\centering
\includegraphics[width=1.0\textwidth]{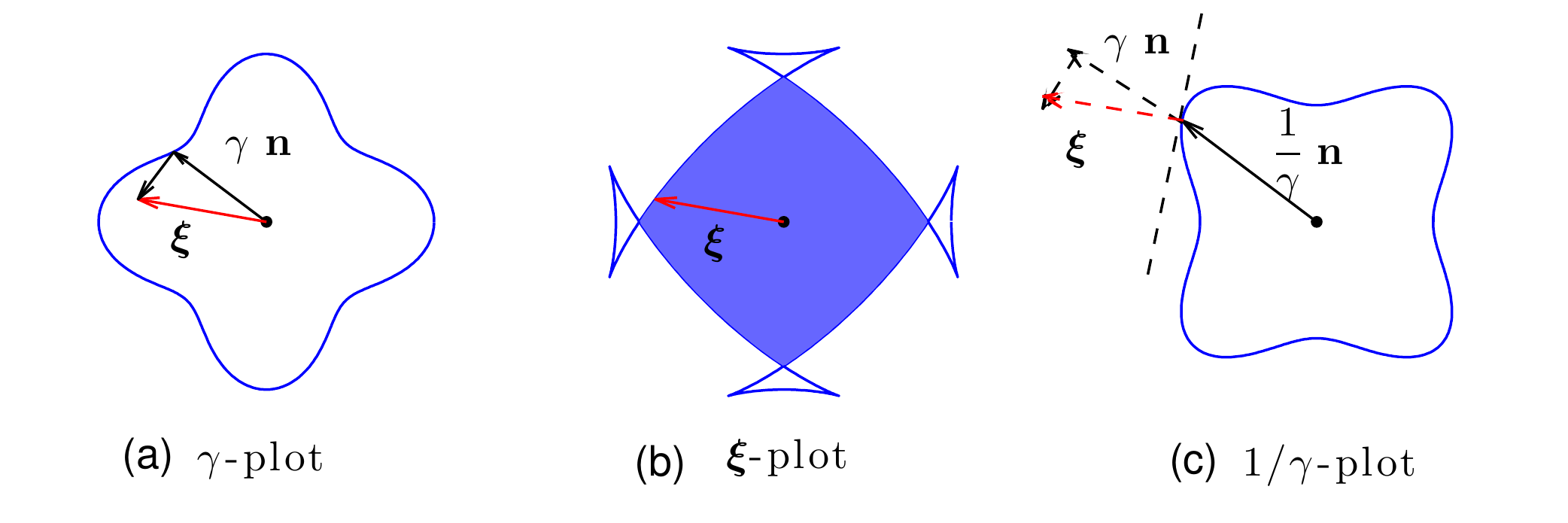}
\caption{A schematic illustration of $\gamma$-plot, $\boldsymbol{\xi}$-plot and $1/\gamma$-plot.
Here, the surface energy density is chosen as a four-fold crystalline symmetry type. It is strongly anisotropic, and its
$\boldsymbol{\xi}$-plot has four ears. The shaded blue region represents its equilibrium shape.}
\label{fig:closedSF}
\end{figure}
Furthermore, in order to effectively and briefly model anisotropic solid materials, Cahn and Hoffman~\cite{Hoffman72, Cahn74} introduced a $\boldsymbol{\xi}$-vector formulation to express the surface energy density of solid materials instead of directly using the traditional scalar function $\gamma$. This vector is mathematically defined as~\cite{Wheeler96,Wheeler99,Eggleston01}
\begin{equation}
\boldsymbol{\xi}(\vec n)=\nabla\hat{\gamma}(\vec p)\Big|_{\vec p=\vec n},
\end{equation}
where $\hat{\gamma}$ is a homogenous extension of the function $\gamma(\vec n)$ from unit vectors to non-zero vectors
$\vec p$ of arbitrary magnitude such that
\begin{equation}
\hat{\gamma}(\vec p) = |\vec p|\gamma(\frac{\vec p}{|\vec p|}),\quad\forall~\vec p\in\mathbb{R}^2\backslash\{\vec 0\}.
\end{equation}
In two dimensions, the surface energy density can also be represented in the form of $\gamma(\theta)$, with $\theta\in[-\pi,\pi]$ representing the orientation angle between the outer normal vector $\vec n$ of the interface and the $y$-axis direction. Under the circumstances, the Cahn-Hoffman $\boldsymbol{\xi}$-vector can be decomposed into two vectors along its local normal and tangential directions (as shown in Fig.~\ref{fig:closedSF}(a)),
\begin{equation}
\boldsymbol{\xi}= \gamma(\theta)\vec n  - \gamma^{\prime}(\theta)\boldsymbol{\tau},
\label{eqn:cahnhoff2}
\end{equation}
where $\boldsymbol{\tau}=(\cos\theta,\sin\theta)^T$ and $\vec n=(-\sin\theta,\cos\theta)^T$ represent the unit tangential and outer normal vectors respectively, and the superscript prime represents the derivative with respect to the variable $\theta$.

For solid thin films during solid-state dewetting process, the equilibrium shape can be mathematically described as that
it minimizes the total free energy $\int_\Gamma\gamma(\theta)\,d\Gamma$ with a prescribed enclosed area/volume
in 2D/3D. The equilibrium shape can be geometrically constructed by the well-known Wulff construction \cite{Wulff01}, and the resulted convex equilibrium shape is often called as Wulff shape. In contrast, the $\boldsymbol{\xi}$-plot (as shown in Fig.~\ref{fig:closedSF}(b)) can be regarded as a mathematical representation of the equilibrium shape.
When the surface stiffness, defined as $\widetilde{\gamma}$, is always positive for all orientations (i.e., the weakly anisotropic case), the $\boldsymbol{\xi}$-plot is just the boundary curve of the Wulff shape up to a scaling. However, when the surface stiffness becomes negative for some orientations (i.e., the strongly anisotropic case and its $1/\gamma$-plot will become non-convex, shown as Fig.~\ref{fig:closedSF}(c)), the $\boldsymbol{\xi}$-plot will have distinct ``ears'', and the Wulff construction tells us that cutting off the ``ears''
gives the equilibrium shape~\cite{Wulff01,Mullins62}, as illustrated in Fig.~\ref{fig:closedSF}(b). In summary,
according to the expressions of surface stiffness, i.e., $\widetilde{\gamma}=(\rm{H}_{\gamma}(\vec n)\boldsymbol{\tau})\cdot\boldsymbol{\tau}=\gamma(\theta)+\gamma^{\prime\prime}(\theta)$,
we can divide the anisotropy into the two cases from mathematics: (a) weakly anisotropic when
$\widetilde{\gamma}>0$ for all orientations; (b) strongly anisotropic when there exist some orientations
for which $\widetilde{\gamma}<0$.
Here, $\rm{H}_\gamma$ is defined as the Hessian matrix of $\hat{\gamma}$, i.e., $\rm{H}_{\gamma}(\vec n)=\nabla^2\hat{\gamma}(\vec p)\Big|_{\vec p=\vec n}$. We note here that the positiveness of the surface stiffness is crucial to the stability of the Wulff shape~\cite{Winklmann06}, well-posedness of the surface diffusion flow \cite{Jiang16} and numerical analysis of this kind of problems \cite{Deckelnick05}.

For the kinetic evolution of solid-state dewetting of thin films, the surface diffusion~\cite{Mullins57} and contact line migration~\cite{Wang15,Zhao17} are recognized as the two main kinetic features. Srolovitz and Safran first proposed a sharp-interface model to investigate the hole growth under the three assumptions, i.e., isotropic surface energy, small slope profile and cylindrical symmetry~\cite{Srolovitz86}. Based on the model, Wong {\it {et al.}} designed a ``marker particle'' numerical method for numerically solving the two-dimensional fully-nonlinear isotropic sharp-interface model
(i.e., without the small slope assumption), and to investigate the two-dimensional edge retraction of a semi-infinite step film~\cite{Wong00}. Dornel {\it {et al.}} designed another numerical scheme to study the pinch-off phenomenon of two-dimensional island films with high aspect ratios during solid-state dewetting~\cite{Dornel06}. Jiang {\it {et al.}}
designed a phase-field model for simulating solid-state dewetting of thin films with isotropic surface energies, and this approach can naturally capture the topological changes that occur during evolution~\cite{Jiang12}. Although most of the above models are focused on the isotropic surface energy case, recent experiments have clearly demonstrated that the kinetic evolution that occurs during solid-state dewetting is strongly affected by crystalline anisotropy~\cite{Thompson12,Leroy16}. In order to investigate surface energy anisotropy effect, many approaches have been proposed and discussed, such as a discrete model~\cite{Dornel06}, a kinetic Monte Carlo model~\cite{Pierre09b,Dufay11}, a crystalline model~\cite{Carter95,Zucker13} and continuum models based on PDEs~\cite{Wang15,Jiang16,Bao17}.
Recently, based on the thermodynamic variation and smooth scalar-field perturbation method, Jiang {\it {et al.}} proposed
a two-dimensional sharp-interface model for investigating solid-state dewetting of thin films with anisotropic surface energies~\cite{Wang15,Jiang16}, and furthermore, they designed a parametric finite element method (PFEM) for solving the sharp-interface model~\cite{Bao17}. However, the smooth scalar-field perturbation method is very complicated and it does not make use of the problem's variational structure, i.e., Cahn-Hoffman $\boldsymbol{\xi}$-vector, and the physical meaning of its variational result is also not direct and clear. Furthermore, this approach is very difficult to generalize to three dimensions.

Therefore, based on the $\boldsymbol{\xi}$-vector, in this paper we present a new approach to deriving the sharp-interface model for solid-state dewetting in two dimensions again~\cite{Wang15,Jiang16}. We claim that this new approach is simpler and very easily extended to the three dimensional case~\cite{Bao18,Zhao17thesis} and its physical meaning is also more direct and clear. The objectives of this paper are: (1) to develop a simpler approach to revisiting the sharp-interface model for simulating solid-state dewetting of thin films via the Cahn-Hoffman $\boldsymbol{\xi}$-vector formulation; (2) to develop a smooth
vector-field perturbation method for calculating the first variation to energy shape functional, which depends on an open curve (or surface) with contact points (or lines); (3) to develop a PFEM for solving the proposed sharp-interface model via the $\boldsymbol{\xi}$-vector, and this approach can deal with any type of surface energy density no matter what its form is ($\gamma(\vec n)$ or $\gamma(\theta)$); (4) to investigate some new physical insights for solid-state dewetting by performing numerical simulations.

The rest of the paper is organized as follows. In Section $2$, we calculate its thermodynamic variation of the total interfacial energy functional with respect to the interface curve $\Gamma$ and the left and right contact points through
the Cahn-Hoffman $\boldsymbol{\xi}$-vector. Subsequently, based on these energetic variations, we derive a kinetic evolution sharp-interface model in Section $3$. In Section $4$, we present its variational formulation of the model and a parametric finite element method is proposed for solving the derived sharp-interface model. In Section $5$, extensive numerical results are performed to demonstrate the high performances of this approach. In Section $6$, we generalize the above sharp-interface model and the numerical scheme to including strongly anisotropic effects. Finally, we draw some conclusions in Section $7$.

\section{Thermodynamic variation via a $\xi$-vector formulation}

\begin{figure}[htp]
\centering
\includegraphics[width=0.7\textwidth]{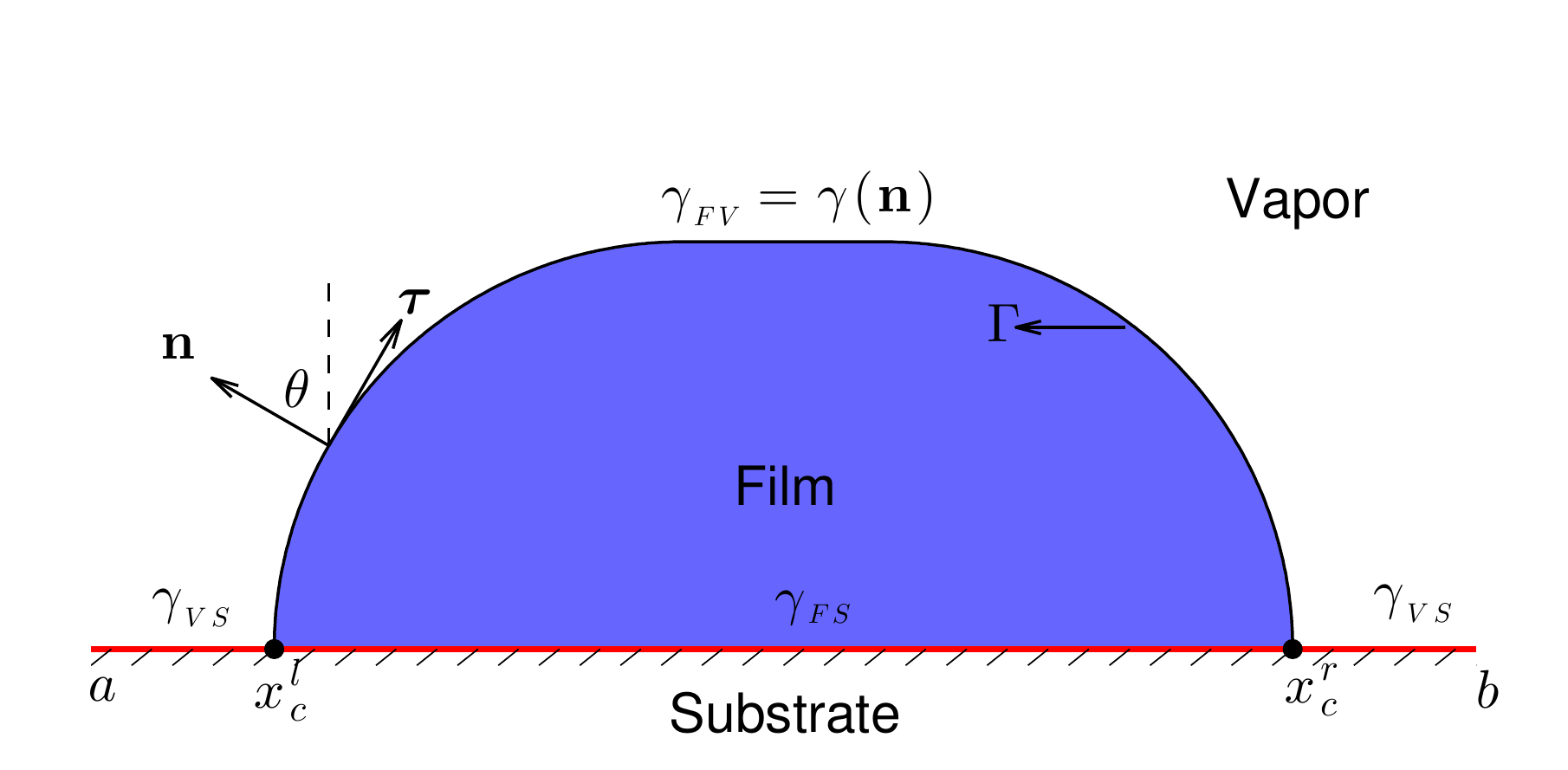}
\caption{A schematic illustration of solid thin film on a rigid, flat substrate ($x$-axis) in two dimensions with two contact points $x_c^l$ and $x_c^r.$}
\label{fig:model}
\end{figure}

\subsection{Interfacial energy functional}

As illustrated in Fig.~\ref{fig:model}, in two dimensions, we consider that an open curve $\Gamma=\vec X(x(s),\;y(s))$, parameterized by the arc length $s$, which separates the vapor and the thin film, attaches on a flat rigid substrate ($x-$axis) with two contact points $x_c^r$ and $x_c^l$. Here $\vec n$ is the unit normal vector that points to the vapor phase and $\boldsymbol{\tau}$ is the unit tangential vector. The total interfacial energy of the system can be written as
\begin{equation}
 W(\Gamma) = \int_{\Gamma_{_{\subFV}}}\gamma_{_{\subFV}}\;d\Gamma_{_{\subFV}} + \int_{\Gamma_{_{\subFS}}}\gamma_{_{\subFS}}\;d\Gamma_{_{\subFS}}+\int_{\Gamma_{_{\subVS}}}\gamma_{_{\subVS}}d\Gamma_{_{\subVS}}.
\end{equation}
Here, $\Gamma_{_{\subFV}}:=\Gamma,\Gamma_{_{\subFS}}$ and $\Gamma_{_{\subVS}}$ represent the film/vapor, film/substrate and vapor/substrate interface, respectively, and $\gamma_{_{\subFV}},\gamma_{_{\subFS}},\gamma_{_{\subVS}}$ represent their corresponding surface energy densities (surface energy per unit length). For solid-state dewetting problems, we often assume that $\gamma_{_{\subFS}},\gamma_{_{\subVS}}$ are two constants, while $\gamma_{_{\subFV}} =\gamma(\vec n)\in C^2(S^1)$ depends on the orientation of the film/vapor interface. By dropping off a constant part, the total interfacial energy can be simplified as the following form (still labeled as $W(\Gamma)$):
\begin{equation}
W(\Gamma) = \int_{\Gamma}\gamma(\vec n)\;ds - (\gamma_{_{\subVS}}-\gamma_{_{\subFS}})(x_c^r-x_c^l),
\label{eqn:surfaceenergy}
\end{equation}
where the first term refers to the film/vapor interfacial energy part, and the second term represents the substrate energy part. In the following, we will introduce a smooth vector-field perturbation method to obtain the first variation of the above energy functional with respect to the open curve $\Gamma$.

\subsection{Smooth vector-field perturbation method and first variation}
First, we introduce an independent parameter $\rho\in I=[0,1]$ to parameterize a family of perturbed curves $\{\Gamma^\epsilon\}_{\epsilon \in [0, \epsilon_0]}$, where the parameter $\epsilon$ controls the amplitude of the perturbation and $\epsilon_0$ is
the maximum perturbation amplitude, i.e.,
\begin{equation}
\Gamma^\epsilon=\vec X(\rho,\epsilon): [0,1]\times[0,\epsilon_0]\rightarrow \mathbb{R}^2,
\label{eqn:parameterization}
\end{equation}
and $\Gamma:= \Gamma^0=\vec X(\rho,0)$. In order to calculate the variation of a shape functional,
we introduce a smooth perturbation vector-field as follows:
\begin{equation}
\vec V(\rho,\epsilon) = \frac{\partial\vec X(\rho,\epsilon)}{\partial\epsilon}, \quad
\forall\,\epsilon \in [0,\epsilon_0],
\label{eqn:Vectorfield}
\end{equation}
and each point on the curve $\Gamma$ is continuously deformed by the above equation defined
by the perturbation vector-field $\vec V$. Note that if the vector-field $\vec V$ is smooth enough,
the family of perturbed curves $\{\Gamma^\epsilon\}$ preserve the regularity of the original curve
$\Gamma$: if $\Gamma$ is of class $C^k$-curves, for any $\epsilon \in [0,\epsilon_0]$,
$\Gamma^\epsilon$ is also of class $C^k$-curves.

Then, assume that given an arbitrary shape functional $F(\Gamma)$, we can define its first variation with respect
to any smooth perturbation vector-field $V$ as
\begin{equation}
\delta F(\Gamma;\vec V) = \lim_{\epsilon \rightarrow 0}\frac{F(\Gamma^\epsilon)-F(\Gamma)}{\epsilon}.
\end{equation}

\begin{lem}\label{le:open curve} Assume that $\Gamma=\vec X(s)\in C^{2}([0,L])\times C^2([0,L])$ is an open smooth curve
with its two endpoints locating at $s=0$ and $s=L$, where $s:=s(\rho), \rho\in [0,1]$ represents the arc length of the curve. If the shape functional $F(\Gamma)=\int_{\Gamma}\gamma(\vec n)\;ds$, then its first variation can be written as:
\begin{equation}\label{eqn:weakopenvariation}
\delta F(\Gamma;\vec V)=-\int_{\Gamma}[({\partial_s\boldsymbol{\xi}})^{\perp}\cdot\vec n]\, (\vec V_0\cdot \vec n)\;ds+ \Bigl[\boldsymbol{\xi}^{\perp}\cdot\vec V_0\Bigr]\Big|_{s=0}^{s=L},
\end{equation}
where $\perp$ represents the clockwise rotation of a vector by 90 degrees, $\boldsymbol{\xi}$ is the Cahn-Hoffman vector and the deformation velocity is denoted as $\vec V_0=\vec V(\rho, 0)$, and $\vec V_0\cdot \vec n$ represents the deformation velocity along the outer normal direction of the interface.

\begin{proof}

We first extend the definition domain of the surface energy density function $\gamma(\vec n)$ from
unit vectors $\vec n$ to arbitrary non-zero vectors $\vec p$ as below
\begin{equation}
\hat{\gamma}(\vec p) = |\vec p|\gamma(\frac{\vec p}{|\vec p|}),\quad\forall\, \vec p\in\mathbb{R}^2\backslash\{{\vec 0}\},
\end{equation}
where $\hat{\gamma}(\vec p)$ extends $\gamma$ as a homogeneous function of the first degree.

The perturbed curve is labeled as $\Gamma^\epsilon:=\vec X(\rho,\epsilon)$. By using the identity $\partial_\rho s=|\partial_\rho\vec X|$, we can have the following expression
\begin{equation}
F(\Gamma^\epsilon)=\int_0^1\gamma(\vec n^\epsilon)|\partial_\rho\vec X(\rho,\epsilon)|\;d\rho.
\end{equation}
Note that the following expressions hold:
\begin{eqnarray}
&&\boldsymbol{\tau}^\epsilon = \frac{\partial_\rho \vec X(\rho,\epsilon)}{\partial_\rho s(\rho,\epsilon)},\qquad \vec n^\epsilon =-(\boldsymbol{\tau^\epsilon})^{\perp}=-\frac{[\partial_\rho\vec X(\rho,\epsilon)]^{\perp}}{|\partial_\rho\vec X(\rho,\epsilon)|},\nonumber\\
&&\nabla\hat{\gamma}(\vec p)\cdot\vec p = \hat{\gamma}(\vec p),\qquad \vec V_0 = \vec V(\rho,0)=\partial_\epsilon\vec X(\rho,\epsilon)\big|_{\epsilon=0}.
\label{eqn:notations}
\end{eqnarray}
Denote $\vec X=\vec X(\rho,0)$, then we can take the Taylor expansion for the following terms at $\epsilon=0$,
\begin{subequations}
\begin{align}
\label{eqn:perb1}
&|\partial_\rho\vec X(\rho,\epsilon)| = |\partial_\rho\vec X| + \frac{\partial_\rho\vec X\cdot\partial_\rho\vec V_0}{|\partial_\rho\vec X|}\epsilon + O(\epsilon^2),\\
\label{eqn:perb2}
&\vec n^\epsilon = \vec n + \Bigr[\frac{\partial_\rho\vec X\cdot\partial_\rho\vec V_0}{|\partial_\rho\vec X|^3}(\partial_\rho\vec X)^\perp-\frac{(\partial_\rho\vec V_0)^\perp}{|\partial_\rho\vec X|}\Bigr]\epsilon + O(\epsilon^2),\\
\label{eqn:perb3}
 &\gamma(\vec n^\epsilon) =\gamma(\vec n) + \nabla\hat{\gamma}(\vec n)\cdot\Bigr[\frac{\partial_\rho\vec X\cdot\partial_\rho\vec V_0}{|\partial_\rho\vec X|^3}(\partial_\rho\vec X)^\perp-\frac{(\partial_\rho\vec V_0)^\perp}{|\partial_\rho\vec X|} \Bigr] \epsilon+ O(\epsilon^2).
\end{align}
\end{subequations}
The first variation of the shape functional can be written as
\begin{equation}
\delta F(\Gamma;\vec V)=\lim_{\epsilon\rightarrow0}\frac{F(\Gamma^\epsilon)-F(\Gamma)}{\epsilon} =\int_0^1\lim_{\epsilon\rightarrow0}\frac{1}{\epsilon}\Bigl[\gamma(\vec n^\epsilon)|\partial_\rho\vec X(\rho,\epsilon)|-\gamma(\vec n)|\partial_\rho\vec X(\rho,0)|\Bigr]\;d\rho.\nonumber
\end{equation}
By substituting Eqs.~\eqref{eqn:perb1} - \eqref{eqn:perb3} into the above equation, and using the identity $\nabla\hat{\gamma}(\vec n)\cdot\vec n = \gamma(\vec n)$, we can obtain
\begin{eqnarray}
\delta F(\Gamma;\vec V)&=& \int_{0}^1\nabla\hat{\gamma}(\vec n)\cdot\Bigr[\frac{\partial_\rho\vec X\cdot\partial_\rho\vec V_0}{|\partial_\rho\vec X|^3}(\partial_\rho\vec X)^\perp-\frac{(\partial_\rho\vec V_0)^\perp}{|\partial_\rho\vec X|} \Bigr] |\partial_\rho\vec X|\;d\rho+\int_0^1\gamma(\vec n)\frac{\partial_\rho\vec X\cdot\partial_\rho\vec V_0}{|\partial_\rho\vec X|}\;d\rho\nonumber\\
&=& -\int_0^1\nabla\hat{\gamma}(\vec n)\cdot (\partial_\rho\vec V_0)^{\perp}\;d\rho = \int_0^1\nabla\hat{\gamma}(\vec n)^\perp\cdot \partial_\rho\vec V_0\;d\rho.
\label{eqn:firstvariation}
\end{eqnarray}
By using integration by parts, and Cahn-Hoffman vector $\boldsymbol{\xi}=\nabla\hat{\gamma}(\vec n)$, we can obtain
\begin{eqnarray}
\delta F(\Gamma;\vec V)&=&-\int_0^1(\partial_\rho\boldsymbol{\xi})^{\perp}\cdot\vec V_0\;d\rho + \Bigl[\boldsymbol{\xi}^{\perp}\cdot\vec V_0\Bigr]\Big|_{\rho=0}^{\rho=1}=-\int_{\Gamma}(\partial_s{\boldsymbol{\xi}})^{\perp}\cdot\vec V_0\;ds+ \Bigl[\boldsymbol{\xi}^{\perp}\cdot\vec V_0\Bigr]\Big|_{s=0}^{s=L}. \label{eqn:form}
\end{eqnarray}
Making use of the expressions: $\partial_s{\boldsymbol{\xi}} \sslash \boldsymbol{\tau}$ and $(\partial_s{\boldsymbol{\xi}})^{\perp} \sslash \vec n$, we can immediately obtain the conclusion from the above Eq.~\eqref{eqn:form}.
\end{proof}
\end{lem}

By using the above Lemma.~\ref{le:open curve}, we can easily obtain the following Theorem about the first variation of the free energy functional \eqref{eqn:surfaceenergy}:

\begin{thm}\label{thm:first}
The first variation of the free energy functional \eqref{eqn:surfaceenergy} used in solid-state dewetting problems
with respect to any smooth deformation field $\vec V$ can be written as:
\begin{equation}
\delta W(\Gamma;\vec V)=-\int_{\Gamma}[(\partial_s{\boldsymbol{\xi}})^{\perp}\cdot \vec n]\,(\vec V_0 \cdot \vec n)\;ds + \Bigl[\bigl(\xi_2-(\gamma_{_{\subVS}}-\gamma_{_{\subFS}})\bigr)\,(\vec V_0 \cdot \vec e_1)\Bigl]\Big|^{s=L}_{s=0},
\label{eqn:DlessEgva}
\end{equation}
where $\boldsymbol{\xi}=(\xi_1,\xi_2)$, ${\vec e_1}=(1,0)$ represents the unit vector along the $x$-coordinate (or the substrate line), and $(\vec V_0 \cdot \vec e_1)\big|^{s=L}_{s=0}$ represents the deformation velocity along the substrate line at two contact points.
\begin{proof}

For solid-state dewetting problems, as shown in Fig.~\ref{fig:model}, the contact points must move along the substrate. Under the assumption that the substrate is flat, the velocity field $\vec V(\rho, 0)$ at the two contact points (i.e., the two endpoints of the curve $\Gamma$) must satisfy the constraints: $\vec V_0(s=0) \sslash {\vec e_1}$ and $\vec V_0(s=L) \sslash {\vec e_1}$.

By using the above Eq.~\eqref{eqn:weakopenvariation}, and $\boldsymbol{\xi}^{\perp}=(\xi_2,-\xi_1)$, we can obtain
\begin{eqnarray*}
\delta W(\Gamma;\vec V)&=&-\int_{\Gamma}[(\partial_s{\boldsymbol{\xi}})^{\perp}\cdot \vec n]\,(\vec V_0 \cdot \vec n)\;ds + \Bigl[(\boldsymbol{\xi}^{\perp}\cdot \vec e_1)\,(\vec V_0 \cdot \vec e_1)\Bigl]\Big|^{s=L}_{s=0}-\Bigl[(\gamma_{_{\subVS}}-\gamma_{_{\subFS}})\,(\vec V_0 \cdot \vec e_1)\Bigr]\Big|^{s=L}_{s=0}, \\
&=&-\int_{\Gamma}[(\partial_s{\boldsymbol{\xi}})^{\perp}\cdot \vec n]\,(\vec V_0 \cdot \vec n)\;ds + \Bigl[\bigl(\xi_2-(\gamma_{_{\subVS}}-\gamma_{_{\subFS}})\bigr)\,(\vec V_0 \cdot \vec e_1)\Bigl]\Big|^{s=L}_{s=0}.
\end{eqnarray*}

\end{proof}
\end{thm}

\section{Sharp-interface model via a $\xi$-vector formulation}

By using the above Theorem 2.1, the first variation of the total energy functional \eqref{eqn:surfaceenergy} with respect to the interface $\Gamma$ and two contact points $x_c^l$ and $x_c^r$ can be written as:
\begin{equation}
\frac{\delta W}{\delta\Gamma}=-(\partial_s\boldsymbol{\xi})^{\perp}\cdot\vec n,\quad
\frac{\delta W}{\delta x_c^l}=-\Bigl(\xi_2|_{s=0} - (\gamma_{_{\subVS}}-\gamma_{_{\subFS}})\Bigr),\quad
\frac{\delta W}{\delta x_c^r}=\xi_2|_{s=L} - (\gamma_{_{\subVS}}-\gamma_{_{\subFS}}).
\label{eqn:weakvariation1}
\end{equation}

From the Gibbs-Thomson relation \cite{Sutton95,Mullins57}, the chemical potential $\mu$ of the system is defined as
\begin{equation}
\mu = \Omega_0\frac{\delta W}{\delta\Gamma},
\end{equation}
where $\Omega_0$ represents the atomic volume of the thin film material. The normal velocity of the interface curve $\Gamma$, labeled as $v_n$, is expressed by the following surface diffusion flow \cite{Mullins57,Cahn74}:
\begin{equation}
\vec j = -\frac{D_s\nu}{k_B\,T_e}\nabla_s\, \mu,\qquad v_n=-\Omega_0 (\nabla_s \cdot \vec j)=\frac{D_s\nu\Omega_0}{k_B\,T_e}\partial_{ss}\mu,
\end{equation}
where $\vec j$ is the mass flux along the interface, $\nabla_s$ is the surface gradient operator, $D_s$ is the surface diffusivity, $\nu$ is the number of diffusing atoms per unit length, and $k_B\,T_e$ is the thermal energy. Furthermore, the motion of the two contact points are given by the energy gradient flow, which is determined by the time-dependent Ginzburg-Landau kinetic equations~\cite{Wang15}, i.e.,
\begin{equation}
\frac{dx_c^l(t)}{dt}=-\eta\frac{\delta W}{\delta x_c^l},\qquad\frac{dx_c^r(t)}{dt}=-\eta\frac{\delta W}{\delta x_c^r},
\end{equation}
with $\eta \in (0, +\infty)$ representing the finite contact point mobility.

We choose the characteristic length scale and characteristic surface energy scale as $h_0$ and $\gamma_{_0}$ respectively, the time scale as $\frac{h_0^4}{B\gamma_0}$ with $B=\frac{D_s\nu\Omega_0^2}{k_B\,T_e}$, and the contact line mobility is scaled by $\frac{B}{h_0^3}$. Then, we can obtain a dimensionless sharp-interface model again~\cite{Wang15,Bao17} for solid-state dewetting via a $\boldsymbol{\xi}$-vector formulation, which can be written as follows (for simplicity, we still use the same notations for the variables):
\begin{eqnarray}
\label{eqn:weak1}
&&\partial_{t}\vec{X}=\partial_{ss}\mu \; \vec{n}, \qquad 0<s<L(t), \qquad t>0, \\
&&\mu=-\left(\partial_s\boldsymbol{\xi}\right)^{\perp}\cdot\vec n,\qquad \boldsymbol{\xi} = \nabla\hat{\gamma}(\vec p)\Big|_{\vec p=\vec n};
\label{eqn:weak2}
\end{eqnarray}
where $\Gamma:=\Gamma(t)=\vec X(s,t)=(x(s,t),y(s,t))$ represents the moving film/vapor interface, $s$ is the arc length or distance along the interface, $t$ is the time, $\vec n=(-\partial_{s}y, \partial_{s}x)$ is the interface outer unit normal vector, $\mu:=\mu(s,t)$ is the chemical potential, $\boldsymbol{\xi}=(\xi_1,\xi_2)$ is the dimensionless Cahn-Hoffman vector (scaled by $\gamma_0$) and  $L:=L(t)$ represents the total length of the moving interface. The initial condition is given as
\begin{equation}
\vec{X}(s,0):=\vec{X}_0(s)=(x(s,0),y(s,0))=(x_0(s),y_0(s)), \qquad 0\le s\le L_0:=L(0),
\label{eqn:init}
\end{equation}
satisfying $y_0(0)=y_0(L_0)=0$ and $x_0(0)<x_0(L_0)$, and the boundary conditions are:
\begin{itemize}
\item[(i)] contact point condition
\begin{equation}\label{eqn:weakBC1}
y(0,t)=0, \qquad y(L,t)=0, \qquad t\ge0;
\end{equation}
\item[(ii)] relaxed contact angle condition
\begin{eqnarray}\label{eqn:weakBC2}
\frac{d x_c^l}{d t}= \eta (\xi_2|_{s=0}-\sigma),\qquad
\frac{d x_c^r}{d t}= -\eta(\xi_2|_{s=L}-\sigma),\qquad t\ge0;
\end{eqnarray}
where the dimensionless material constant $\sigma = \frac{\gamma_{_{\subVS}}-\gamma_{_{\subFS}}}{\gamma_0}$,
and $\gamma_0$ is the dimensionless unit of surface energy density.
\item[(iii)] zero-mass flux condition
\begin{equation}
\partial_s \mu(0,t)=0, \qquad \partial_s \mu(L,t)=0,\qquad t\ge0.
\label{eqn:weakBC3}
\end{equation}
\end{itemize}
For the above boundary conditions, condition (i) ensures that the contact points always move along the substrate, condition (ii) allows for the relaxation of the contact angle, and condition (iii) ensures that the total area/mass of the thin film is conserved, implying that there is no mass flux at the contact points. We note here that the above
governing equations \eqref{eqn:weak1}-\eqref{eqn:weak2} are mathematically well-posed when the surface energy is isotropic or weakly anisotropic; if the surface energy is strongly anisotropic, the above equations will become anti-diffusion type, and they are ill-posed. In the strongly anisotropic case, we need to regularize these equations by adding some high-order terms, and we will discuss this case in Section $6$.

\begin{prop}[Mass conservation and energy dissipation]
\label{prop:massenergy}
 Assume that $\Gamma(t)=\vec X(s,t)$ is the solution of Eqs.~\eqref{eqn:weak1}-\eqref{eqn:weak2} coupled with the boundary conditions~\eqref{eqn:weakBC1}-\eqref{eqn:weakBC3},  then the total area/mass of the thin
 film is conserved during the evolution, i.e.,
\begin{equation}\label{eqn:totalmassw}
A(t)\equiv A(0)=\int_{\Gamma(0)}y_0(s)\partial_s x_0(s)\;ds, \qquad
t\ge0,
\end{equation}
and the total free energy of the system decreases during the evolution, i.e.,
\begin{equation}\label{eqn:totalenegw}
W(t)\le W(t_1)\le W(0)=\int_{\Gamma(0)}\gamma(\vec n)\;ds-(x_c^r(0)-x_c^l(0))\sigma, \qquad
 t\ge t_1\ge0.
\end{equation}

\begin{proof}
By directly calculating the time derivative of $A(t)$, we can obtain the following expressions:
\begin{eqnarray*}
    \frac{d}{dt}A(t)
    &=&\frac{d}{dt}\int_{\Gamma(t)}y \partial_s x\;ds
    =\frac{d}{dt}\int_0^1 y \partial_\rho x\;d\rho
    =\int_0^1 (\partial_t y \partial_\rho x+y \partial_{\rho t} x)\;d\rho\nonumber\\
    &=&\int_0^1 (\partial_t y \partial_\rho x-\partial_\rho y \partial_t x)~d\rho + \bigl(y \partial_t x\bigr)\Big|_{\rho=0}^{\rho=1}
    =\int_0^1 (\partial_t x, \partial_t y)\cdot(-\partial_\rho y, \partial_\rho x)\;d\rho
    =\int_{\Gamma(t)}\partial_t \vec{X}\cdot\vec{n}\;ds\nonumber\\
    &=&\int_{\Gamma(t)}\partial_{ss}\mu\;ds=\bigl(\partial_s \mu)\Big|_{s=0}^{s=L(t)}=0,
\end{eqnarray*}
where the last equality follows from the zero-mass flux boundary condition, and it indicates that the
total area/mass is conserved.

To obtain the time derivative of $W(t)$, by making use of Eq.~\eqref{eqn:DlessEgva}, but replacing the perturbation variable $\epsilon$ with the time variable $t$, we can immediately obtain:
\begin{equation*}
\frac{d}{dt}W(t)=-\int_{\Gamma(t)}(\partial_s{\boldsymbol{\xi}})^{\perp}\cdot\partial_t \vec X\; ds + ({\vec \xi_2}|_{s=L(t)}-\sigma)\frac{dx_c^r}{dt}
-({\vec \xi_2}|_{s=0}-\sigma)\frac{dx_c^l}{dt},
\end{equation*}
where we note that $\frac{dx_c^r}{dt}=(\vec V_0 \cdot \vec e_1)\Big |_{s=L(t)}$ and $\frac{dx_c^l}{dt}=(\vec V_0 \cdot \vec e_1)\Big |_{s=0}$.

By using \eqref{eqn:weak1} and \eqref{eqn:weak2} (i.e., the relaxed contact angle and zero-mass flux conditions), we have
\begin{eqnarray*}
\frac{d}{dt}W(t)
=\int_{\Gamma(t)}\mu\partial_{ss}\mu\;ds-\frac{1}{\eta}\Bigl[\Bigl(\frac{dx_c^l}{dt}\Bigr)^2+
\Bigl(\frac{dx_c^r}{dt}\Bigr)^2\Bigr]
=-\int_{\Gamma(t)}(\partial_s\mu)^2\;ds-\frac{1}{\eta}\Bigl[\Bigl(\frac{dx_c^l}{dt}\Bigr)^2+
\Bigl(\frac{dx_c^r}{dt}\Bigr)^2\Bigr]\le0.
\end{eqnarray*}
The last inequality immediately implies the energy dissipation.
\end{proof}
\end{prop}

\section{A parametric finite element method}

In this section, motivated by the {\sl parametric finite element method (PFEM)} recently used for solving a class of geometric partial differential equations (e.g., \cite{Deckelnick05,Bansch05,Barrett07Iso,Barrett07Aniso,Bao17}), we propose a parametric finite element numerical scheme for solving the above proposed sharp-interface mathematical model, i.e., Eqs.~\eqref{eqn:weak1}-\eqref{eqn:weak2} coupled with the boundary conditions Eqs.~\eqref{eqn:weakBC1}-\eqref{eqn:weakBC3}.

\subsection{Variational formulation}

Following the previous notions, we assume that $\Gamma(t)$ is a family of open evolution curves in the plane which intersect with the substrate line ($x$-axis) at the two contact points, where $t\in[0,T]$ represents
the time, then we can parameterize the curves as
\begin{equation}
\Gamma(t)=\vec X(\rho, t): I\times [0,T]\rightarrow \mathbb{R}^2,
\end{equation}
where the time-independent spatial variable $\rho \in I$, and $I$ denotes a fixed reference spatial domain.
For simplicity, we choose it as $I :=[0,1]$.

In order to briefly present its variational formulation of the sharp-interface model,
we introduce the following $L^2$ inner product which depends on the evolution curve $\Gamma(t)$ as
\begin{equation}
\big<u,v\big>_{\Gamma}:=
\int_{I}u(\rho)\cdot v(\rho)|\partial_\rho\vec{X}(\rho,t)|\,d\rho,
\end{equation}
where $u,v\in L^2(I)$ are any scalar (or vector) functions. In addition, here we always assume that $\partial_\rho s(\rho,t)=|\partial_\rho\vec{X}(\rho,t)|\in L^{\infty}(I)$. On the other hand, when the interface curve $\Gamma(t)$ evolves, the $x$-coordinates of two contact points at which $\Gamma(t)$ intersects with the $x$-axis will evolve
according to the relaxed contact angle condition Eq.~\eqref{eqn:weakBC2}, and therefore,
we can define the following Dirichlet-type functional space of the solutions for the proposed sharp-interface
model as
\begin{equation}
H_{a,b}^1(I)=\{u \in H^{1}(I): u(0)=a, u(1)=b\},
\end{equation}
where $a$ and $b$ are two preassigned constants which are related to $x$-coordinates (or $y$-coordinates) of
the two contact points at a fixed time, respectively. For simplicity, we denote $H_0^1(I):=H_{0,0}^1(I)$.

Now, we can define the variational formulation of the above sharp-interface model for simulating the solid-state dewetting of thin films: given an initial curve $\Gamma(0)=\vec X(\rho,0)=\vec X_0(s)$ with $s=L_0\rho$
for $\rho \in I$ (defined in Eq.~\eqref{eqn:init}), for any time $t \in (0,T]$, find the evolution curves
$\Gamma(t)=\vec X(\rho,t)\in H_{a,b}^1(I)\times H_{0}^{1}(I)$ with the
$x$-coordinate positions of moving contact points $a=x_c^l(t)\leq x_c^r(t)=b$,
the chemical potential $\mu(\rho,t)\in H^1(I)$ such that
\begin{eqnarray}
&&\big<\partial_{t}\vec{X},~\varphi\vec{n}\big>_{\Gamma}+\big<\partial_{s}\mu,~\partial_{s}\varphi\big>
_{\Gamma}=0, \quad \forall\; \varphi\in H^{1}(I), \label{eqn:weakvf1}\\[1em]
&&\big<\mu\vec n,~\boldsymbol{\omega}\big>_{\Gamma}-\big<\boldsymbol{\xi}^{\perp},~\partial_s\boldsymbol{\omega}\big>_{\Gamma}=0, \quad\forall\; \boldsymbol{\omega}\in H^{1}_0(I)\times H^1_0(I), \label{eqn:weakvf2}
\end{eqnarray}
coupled with that the positions of the moving contact points, i.e., $x_c^l(t)$ and $x_c^r(t)$,
are updated by the relaxed contact angle boundary condition, i.e., Eq.~\eqref{eqn:weakBC2}.
Here, the Cahn-Hoffman $\boldsymbol{\xi}$-vector is determined from the surface energy density
$\gamma(\vec n)$ and the curve orientation $\vec n$,
i.e., $\boldsymbol{\xi} = \nabla\hat{\gamma}(\vec p)\Big|_{\vec p=\vec n}$.
It is noted that Eq.~\eqref{eqn:weakvf1} is obtained by reformulating Eq.~\eqref{eqn:weak1} as $\partial_{t}\vec{X} \cdot \vec{n}=\partial_{ss}\mu$, then multiplying a scalar test function $\varphi$ on the both sides
and integrating over the interface curve $\Gamma(t)$, and finally using the integration by parts and
the zero-mass flux boundary condition \eqref{eqn:weakBC3}. Similarly, Eq.~\eqref{eqn:weakvf2} is obtained
by reformulating Eq.~\eqref{eqn:weak2} as $\mu\;\vec n = -\partial_s \boldsymbol{\xi}^\perp$, multiplying a vector- valued test function $\boldsymbol{\omega}$ on its both sides, and the integration by parts.

\subsection{Fully-discrete scheme}

A uniform partition of $I$ is given as:
$\rho \in I=[0,1]=\bigcup_{j=1}^{N}I_j=\bigcup_{j=1}^{N}[\rho_{j-1},\rho_{j}]$, where $N$ denotes the number of divided small intervals, and $\rho_{j}=jh$ denotes the interval nodes with the uniform interval length $h=1/N$. In addition, we subdivide the time interval as $0=t_{0}<t_{1}<\ldots<t_{M-1}<t_{M}=T$ with $\tau_m = t_{m+1}-t_m$. Define the finite dimensional spaces to approximate $H^1(I)$ and $H_{a,b}^1(I)$  as
\begin{eqnarray}
\label{eqn:FEMspace1}
&&V^h:=\{u\in C(I):\;u\mid_{I_{j}}\in P_1,\quad \forall \, j=1,2,\ldots,N\}\subseteq H^1(I),\\
\label{eqn:FEMspace2}
&&\mathcal{V}^h_{a,b}:=\{u \in V^h:\;u(0)=a,~u(1)=b\}\subseteq H^1_{a,b}(I),
\end{eqnarray}
where $a$ and $b$ are two given constants, $P_1$ denotes the polynomial with degrees at most $1$. And again,  for simplicity, we denote $\mathcal{V}^h_0=\mathcal{V}^h_{0,0}$.

Since we use the $P_1$ (linear) elements to approximate the moving curves, the numerical solutions for the moving interfaces are polygonal curves. If we introduce that, $\vec h_j^m:=\vec X^m(\rho_{j})-\vec X^m(\rho_{j-1})$, is a straight line (or a vector) which connects with the marker points $\vec X^m(\rho_{j})$ and $\vec X^m(\rho_{j-1})$, where $j=1\rightarrow N$, then we can denote the evolution curve at time $t=t_m$ as: $\vec X^m=\bigcup_{j=1}^N \vec h_j^m$, and its tangential, normal vector of the numerical solution $\Gamma^m$ are step functions with possible discontinuities or jumps at nodes $\rho_j$. For two piecewise continuous scalar or vector functions $u$ and $v$
defined on the interval $I$, with possible jumps at the nodes $\{\rho_j\}_{j=1}^{N-1}$,
we can define the mass lumped inner product $\big<\cdot,\cdot\big>_{\Gamma^m}^h$ over $\Gamma^m$ as
\begin{equation}
\big<u,~v\big>_{\Gamma^m}^h:=\frac{1}{2}\sum_{j=1}^{N}\Big|\vec{X}^m(\rho_{j})-
\vec{X}^m(\rho_{j-1})\Big|\Big[\big(u\cdot v\big)(\rho_j^-)+\big(u\cdot v\big)(\rho_{j-1}^+)\Big],
\end{equation}
where $u(\rho_j^\pm)=\lim\limits_{\rho\to \rho_j^\pm} u(\rho)$. The unit tangential vector can be computed as $\boldsymbol{\tau}^m_j=\boldsymbol{\tau}^m|_{I_j}=\frac{\vec h_j^m}{|\vec h_j^m|}$, and the normal vector can be numerically computed as $\vec{n}^m=-({\boldsymbol{\tau}}^m)^{\perp}$.

Let $\Gamma^{m}:=\vec{X}^{m}$, $\vec{n}^m$ and $\mu^m$ be the numerical approximations of the moving curve $\Gamma(t_m):=\vec{X}(\cdot,t_m)$, the normal vector $\vec{n}$ and the chemical potential $\mu$ at time $t_{m}$, respectively. For simplicity, we denote $\vec X^m(\rho_j)=(x_j^m,y_j^m)$.
Take $\Gamma^0=\vec{X}^0\in \mathcal{V}^h_{x_0(0),x_0(L_0)} \times \mathcal{V}^h_0$
such that $\vec{X}^0(\rho_j)=\vec{X}_0(s_j^0)$ with $s_j^0=jL_0/N=L_0\rho_j$ for $j=0,1,\ldots,N$,
then a semi-implicit {\sl parametric finite element method (PFEM)} for the variational problem
\eqref{eqn:weakvf1}-\eqref{eqn:weakvf2} can be given as: for $m\ge0$,
find $\Gamma^{m+1}=\vec{X}^{m+1}\in \mathcal{V}^h_{a,b}\times
\mathcal{V}^h_0$ with the $x$-coordinate positions of the
moving contact points $a:=x_c^l(t_{m+1})\leq b:=x_c^r(t_{m+1})$ and $\mu^{m+1}\in V^h$ such that
\begin{subequations}\label{eq_ch3:weq}
    \begin{align}
         &\big<\frac{\vec X^{m+1}-\vec X^m}{\tau_m},~\varphi_h\vec{n}^m\big>_{\Gamma^m}^h+\big<\partial_{s}\mu^{m+1},~\partial_{s}\varphi_h\big>
_{\Gamma^m}^h=0, \quad \forall\; \varphi_h\in V^h, \label{eqn:weakfull1}\\
        &\big<\mu^{m+1}\vec n^m,~\boldsymbol{\omega_h}\big>_{\Gamma^m}^h-\big<(\boldsymbol{\xi}^{m+\frac{1}{2}})^{\perp},~\partial_s\boldsymbol{\omega_h}\big>_{\Gamma^m}^h=0,\quad\forall\;\boldsymbol{\omega}_h\in \mathcal{V}^h_0\times \mathcal{V}^h_0,\label{eqn:weakfull2}
    \end{align}
\end{subequations}
where the $x$-coordinates of the two contact point positions $x_c^l(t_{m+1})$ and $x_c^r(t_{m+1})$ are updated via the relaxed contact angle condition Eq.~\eqref{eqn:weakBC2} by using the forward Euler numerical approximation,
\begin{equation}
\label{eqn:weakfull3}
\frac{x_c^l(t_{m+1}) - x_c^l(t_m)}{\tau_m} =\eta\Bigl[\xi_2^m\Big|_{\rho=0} - \sigma \Bigr],\quad\frac{x_c^r(t_{m+1}) - x_c^r(t_m)}{\tau_m} = -\eta\Bigl[\xi_2^m\Big|_{\rho=1} - \sigma \Bigr].
\end{equation}
According to the specific form of the surface energy density, we can define the numerical approximation
term $\boldsymbol{\xi}^{m+\frac{1}{2}}$ in the above scheme as follows: If the surface energy density is expressed as the form of $\gamma(\vec n)$, the numerical approximation
term $\boldsymbol{\xi}^{m+\frac{1}{2}}$ can be defined as:
\begin{equation}
\boldsymbol{\xi}^{m+\frac{1}{2}}=\gamma(\vec n^m)\vec n^{m+1}+(\boldsymbol{\xi}^{m}\cdot \boldsymbol{\tau}^{m})\boldsymbol{\tau}^{m+1}.
\label{eqn:gamman}
\end{equation}
The main idea comes from that the Cahn-Hoffman vector can be decomposed as: $\boldsymbol{\xi}=
\gamma(\vec n)\vec n + (\boldsymbol{\xi}\cdot \boldsymbol{\tau})\boldsymbol{\tau}$, where $\boldsymbol{\xi}\cdot\vec n=\gamma(\vec n)$, and then we use the semi-implicit discretization. Furthermore, in some literatures, the surface energy density function $\gamma(\vec n)$ can be chosen as a special form, e.g., Riemannian metric form~\cite{Barrett07Aniso,Deckelnick05}
\begin{equation}
\gamma(\vec n) = \sum_{k=1}^K\sqrt{G_k\vec n\cdot\vec n},\qquad \boldsymbol{\xi}(\vec n) = \sum_{k=1}^K\Bigl[\sqrt{G_k\vec n\cdot\vec n}\Bigr]^{-1} G_k\vec n,
\label{eqn:riemann}
\end{equation}
where $G_k$, $k=1,\cdots,K$, is a symmetric positive definite matrix. In this special case, the numerical approximation
term $\boldsymbol{\xi}^{m+\frac{1}{2}}$ can be defined as:
\begin{equation}
\boldsymbol{\xi}^{m+\frac{1}{2}} = \sum_{k=1}^K \Bigl[\sqrt{G_k\vec n^m\cdot\vec n^m}\Bigr]^{-1}G_k\vec n^{m+1}.
\end{equation}

On the other hand, in two dimensions, the surface energy density function $\gamma(\vec n)$ can be equivalently expressed as the form of $\gamma(\theta)$, where $\vec n=(\cos\theta, \sin\theta)$, $\theta \in [-\pi,\pi]$ represents the local orientation (i.e., the angle between the interface outer normal $\vec n$ and y-axis), and by simple calculations we can
evaluate the linear approximation $\boldsymbol{\xi}^{m+\frac{1}{2}}$ in Eq.~\eqref{eqn:weakfull2}
to the Cahn-Hoffman vector as
\begin{equation}
\boldsymbol{\xi}^{m+\frac{1}{2}}=\gamma(\theta^m)\vec n^{m+1}-\gamma\,'(\theta^m)\boldsymbol{\tau}^{m+1},
\label{eqn:discahn}
\end{equation}
where the value of the orientation angle $\theta$ is explicitly calculated at time $t=t^m$, and we have used
$\boldsymbol{\xi}\cdot \boldsymbol{\tau}=-\gamma\,'(\theta)$ in Eq.~\eqref{eqn:gamman}.

For the above semi-implicit parametric finite element scheme, we have the following theorem:

\begin{thm}[Well-posedness for the PFEM scheme] The above discrete variational problem Eqs.~\eqref{eqn:weakfull1}-\eqref{eqn:weakfull2}
admits a unique solution (i.e., well-posed).

\begin{proof}
Note that the two moving contact points is first updated explicitly according to the relaxed boundary condition,
and therefore, the boundary conditions of the variables for the above discrete variational problem can be regarded as the Dirichlet type. Proving that the resulted linear system has a unique solution is equivalent to proving that the corresponding homogeneous linear system has zero solutions, i.e., the system can be reduced to: find $\{\vec X^{m+1},\mu^{m+1}\}\in\{\mathcal{V}_0^h\times\mathcal{V}_0^h, V^h\}$ such that
\begin{subequations}
    \begin{align}
         &\big<\vec X^{m+1},~\varphi_h\vec{n}^m\big>_{\Gamma^m}^h+\big<\partial_{s}\mu^{m+1},~\partial_{s}\varphi_h\big>
_{\Gamma^m}^h=0, \quad \forall\; \varphi_h\in V^h, \label{eqn:weakhome1}\\
        &\big<\mu^{m+1}\vec n^m,~\boldsymbol{\omega_h}\big>_{\Gamma^m}^h-\big<(\boldsymbol{\xi}^{m+\frac{1}{2}})^\perp,~\partial_s\boldsymbol{\omega_h}\big>_{\Gamma^m}^h=0,\quad\forall\;\boldsymbol{\omega}_h\in \mathcal{V}^h_0\times \mathcal{V}^h_0,\label{eqn:weakhome2}
    \end{align}
\end{subequations}
with $\boldsymbol{\xi}^{m+\frac{1}{2}} = \gamma(\theta^m)(-\partial_s\vec X^{m+1})^\perp-\gamma'(\theta^m)\partial_s\vec X^{m+1}$ defined in Eq.~\eqref{eqn:discahn}. Now if we set $\varphi_h = \mu^{m+1},\boldsymbol{\omega}_h =\vec X^{m+1}$ and by noting that $(\partial_s\vec X^{m+1})^\perp\cdot\partial_s\vec X^{m+1}=0$, we can obtain
\begin{equation}
\Big<\partial_s\mu^{m+1},~\partial_s\mu^{m+1}\Big>_{\Gamma^m}^h+ \Big<\gamma(\theta^m)\partial_s\vec X^{m+1},~\partial_s\vec X^{m+1}\Big>_{\Gamma^m}^h=0.
\end{equation}
Because the surface energy density $\gamma(\theta)$ is always non-negative, the above equation directly tells us that $\vec X^{m+1}=\vec 0$, $\mu^{m+1}=\vec 0$. Therefore, the corresponding homogeneous linear system only has zero solutions, which indicates the discrete scheme has a unique solution.
\end{proof}
\end{thm}
Note that in the above proof, we assume that the surface energy density is of the $\gamma(\theta)$ form; if it is written
as the $\gamma(\vec n)$ form, the proof is the same. On the other hand, if the surface energy density is specially chosen as the Riemannian metric form, the proof can be found in \cite{Barrett07Aniso}.

The above proposed PFEM scheme via the Cahn-Hoffman $\boldsymbol{\xi}$ has many advantages over the PFEM scheme previously proposed by us in~\cite{Bao17}: (a) first, in the present PFEM, we only need to solve a linear algebra system which includes the unknown variables $\{\vec X^{m+1},\mu^{m+1}\}$, while it includes $\{\vec X^{m+1},\mu^{m+1},\kappa^{m+1}\}$ in the previous one; (b) second, the present PFEM is well-posed and can work for any form of the surface energy density function, while the previous one only works for the $\gamma(\theta)$ form and can be
proved to be well-posed only in the isotropic surface energy case; (c) third, the present scheme only needs to deal with
the first derivative of $\gamma(\theta)$, while the previous needs to compute its second derivative.
In addition, we note here that in practical simulations, when the surface energy anisotropy becomes stronger and stronger, the interface curve will form sharper and sharper corners. Under these circumstances, we need to redistribute mesh points at evenly spaced arc lengths, and a kind of mesh redistribution algorithm which can conserves the total area can be found in the reference~\cite{Bansch05}.

For any form of $\gamma(\vec n)$, how to prove that the above fully-discrete scheme Eqs.~\eqref{eqn:weakfull1}-\eqref{eqn:weakfull2} preserves the discrete energy-dissipation
property seems difficult. But in some special forms, for example, if $\gamma(\vec n)$ is chosen as the Riemannian metric form, i.e.,~Eq.\eqref{eqn:riemann}, Barrett {\it et al.}~\cite{Barrett07Aniso} can prove a stronger conclusion that the scheme is unconditionally energy-stable (i.e., energy-dissipative regardless of how to choose $\tau_m$ and $h$) for closed evolution curves with periodical boundary conditions.
Some more generalized PFEM schemes which can ensure the stability and are applicable for all types of anisotropy have been discussed and developed in the reference~\cite{Deckelnick05}.  The main idea behind these schemes is to explicitly evaluate the nonlinear term $\boldsymbol{\xi}$ by adding a stabilized term to Eq.~\eqref{eqn:weakhome2} on the right hand side. The fully-discrete stabilized PFEM scheme for solving the sharp-interface model can be written as: for $m\ge0$,
find $\Gamma^{m+1}=\vec{X}^{m+1}\in \mathcal{V}^h_{a,b}\times
\mathcal{V}^h_0$ with the $x$-coordinate positions of the
moving contact points $a:=x_c^l(t_{m+1})\leq b:=x_c^r(t_{m+1})$ and $\mu^{m+1}\in V^h$ such that
{\small
\begin{subequations}
    \begin{align}
         &\Big<\frac{\vec X^{m+1}-\vec X^m}{\tau_m},~\varphi_h\vec{n}^m\Big>_{\Gamma^m}^h+\Big<\partial_{s}\mu^{m+1},~\partial_{s}\varphi_h\Big>
_{\Gamma^m}^h=0, \quad \forall\; \varphi_h\in V^h, \label{eqn:weakfull5}\\
        &\big<\mu^{m+1}\vec n^m,~\boldsymbol{\omega_h}\big>_{\Gamma^m}^h-\Big<(\boldsymbol{\xi}^{m})^{\perp},~\partial_s\boldsymbol{\omega_h}\Big>_{\Gamma^m}^h=
        \lambda\Big<\gamma(\vec n^m)\,\partial_s(\vec X^{m+1}-\vec X^m),~\partial_s\boldsymbol{\omega}_h\Big>_{\Gamma^m}^h,\quad\forall\;\boldsymbol{\omega}_h\in \mathcal{V}^h_0\times \mathcal{V}^h_0,\label{eqn:weakfull6}
    \end{align}
\end{subequations}}
where the two contact point positions $x_c^l(t_{m+1})$ and $x_c^r(t_{m+1})$ are first determined by Eqs.~\eqref{eqn:weakfull3} via forward Euler scheme. Here, $\lambda$ is a stabilized parameter (often chosen as a positive constant). This stabilized PFEM is also a good candidate numerical method for solving our proposed model.
But, since the stabilized term may influence the accuracy of the scheme, especially when
the stabilized parameter $\lambda$ is chosen to be large to control the stability, so in real simulations
this scheme is not our first option compared to the former PFEM scheme, i.e., Eqs.~\eqref{eqn:weakfull1}-\eqref{eqn:weakfull2}.

\section{Numerical results}

\begin{figure}[!htp]
\centering
\includegraphics[width=0.80\textwidth]{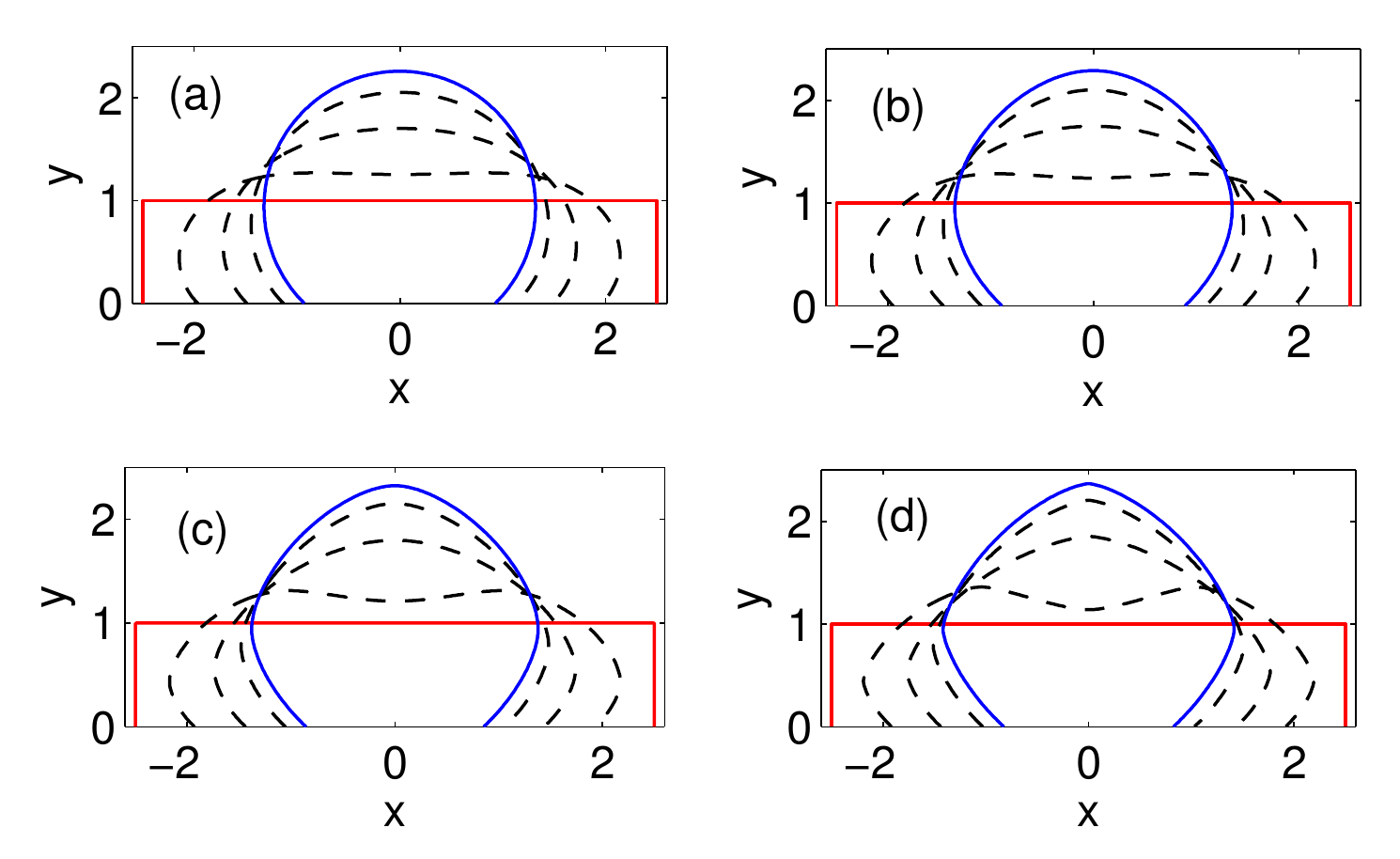}
\caption{\label{fig:wkmbeta1} Several steps in the evolution of small, initial rectangle islands (shown in red) towards their equilibrium shapes (shown in blue) for different degrees of the anisotropy $\beta$ with the crystalline symmetry order $k=4$, where the parameters are chosen as $\sigma=\cos\frac{3\pi}{4}$, and the degree of the anisotropy: (a) $\beta=0$, (b) $\beta=0.02$, (c) $\beta=0.04$, (d) $\beta=0.06$.}
\end{figure}

Based on the sharp-interface model and numerical methods presented above, we will perform numerical simulations in this section for investigating solid-state dewetting of thin films with different kinds of surface energy anisotropies in several different initial geometries. The numerical simulations are performed by using the PFEM \eqref{eqn:weakfull1}-\eqref{eqn:weakfull2}, except where noted. The contact point mobility $\eta$ determines the relaxation rate of the dynamical contact angle to the equilibrium contact angle, and in principle, it is a material parameter which should be determined either from physical experiments or microscopic (e.g., molecular dynamical) simulations. In the following numerical simulations, we always choose a large contact point mobility $\eta=100$. The choice of a large contact point mobility (e.g., $\eta=100$) tends to drive the contact angle very quickly to converge to the equilibrium contact angle, resulting in an equilibrium shape that effectively minimizes the total surface energy with the equilibrium contact angles. The influence of the parameter $\eta$ on the solid-state dewetting evolution process and equilibrium shapes has been discussed in~\cite{Wang15}.

\subsection{Small islands}

The first type of anisotropic surface energy density we will investigate is the $k$-fold smooth crystalline surface energy, which is usually defined as the $\gamma(\theta)$ form
 \begin{equation}
 \gamma(\theta) = 1 + \beta\cos(k\theta), \quad \theta\in[-\pi,\pi],
 \label{eqn:kfoldenergy}
 \end{equation}
where $\beta\geq 0$ controls the degree of the anisotropy, $k$ is the order of the rotational symmetry (usually taken as $k=2,3,4,6$ for crystalline materials). For this surface energy, when
$\beta=0$, it is isotropic; when $0<\beta<\frac{1}{k^2-1}$, it is weakly anisotropic;
and when $\beta>\frac{1}{k^2-1}$, it is strongly anisotropic.

\begin{figure}
\centering
\includegraphics[width=0.85\textwidth]{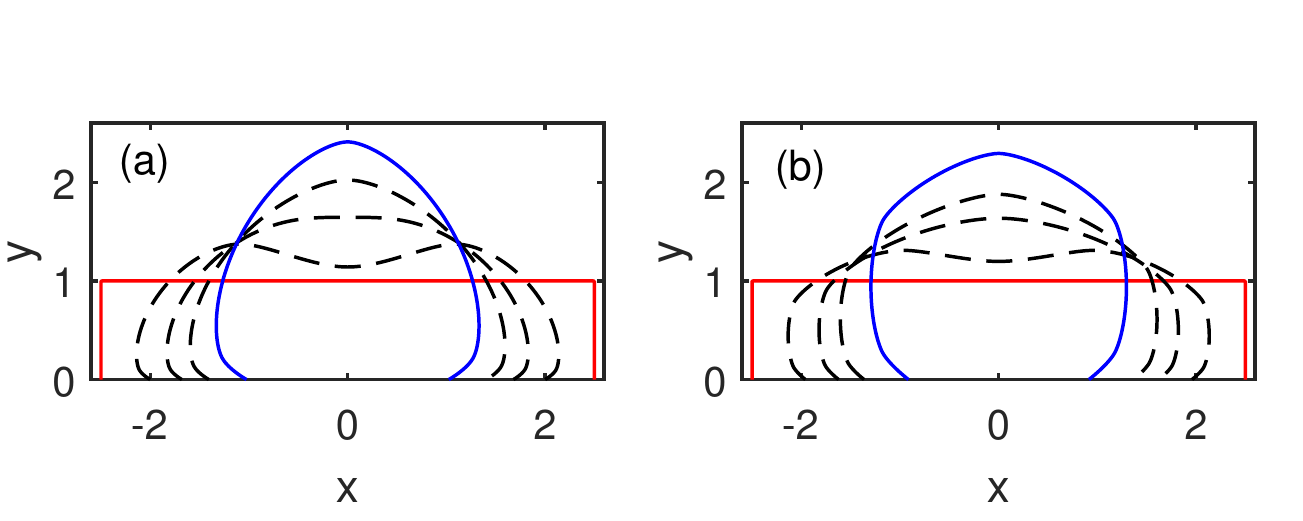}
\caption{\label{fig:wkmbeta2} Several steps in the evolution of small initial rectangle islands (shown in red) towards their equilibrium shapes (shown in blue), where the parameters are chosen as $\sigma=\cos\frac{3\pi}{4}$, (a) $k=3,\beta=0.1$; (b) $k=6,\beta=0.022$.}
\end{figure}

We start with the numerical examples for the $4$-fold anisotropy under four different degrees $\beta$.  As depicted in Fig.~\ref{fig:wkmbeta1}, it shows the evolution process of an initially rectangular thin film (in red solid lines) towards its equilibrium shapes (in blue solid lines). The initial rectangular island film is chosen as the length $5$ and height $1$. From Fig.~\ref{fig:wkmbeta1}(a) to Fig.~\ref{fig:wkmbeta1}(d), the material constants are all chosen $\sigma=\cos(\frac{3\pi}{4})$, while the degree of the anisotropy $\beta$ increases gradually from $\beta=0$ to $\beta=0.06$. The number of the grid points is chosen as $N=400$, and the time step is chosen as a fixed value $\tau=2\times 10^{-4}$. As can be seen clearly in Fig.~\ref{fig:wkmbeta1}, the equilibrium shapes (in blue solid lines) gradually change from a circular arc to an anisotropic shape with increasingly sharper and sharper corners, and the number of ``facets'' in the equilibrium shape also exhibits the $4$-fold geometric symmetry. Moreover, we also test the numerical examples by choosing different symmetry orders $k$, and observe that the $k$-fold symmetry appears in the equilibrium shape (as shown in Fig.~\ref{fig:wkmbeta2}).

\begin{figure}[htp!]
\centering
\includegraphics[width=0.80\textwidth]{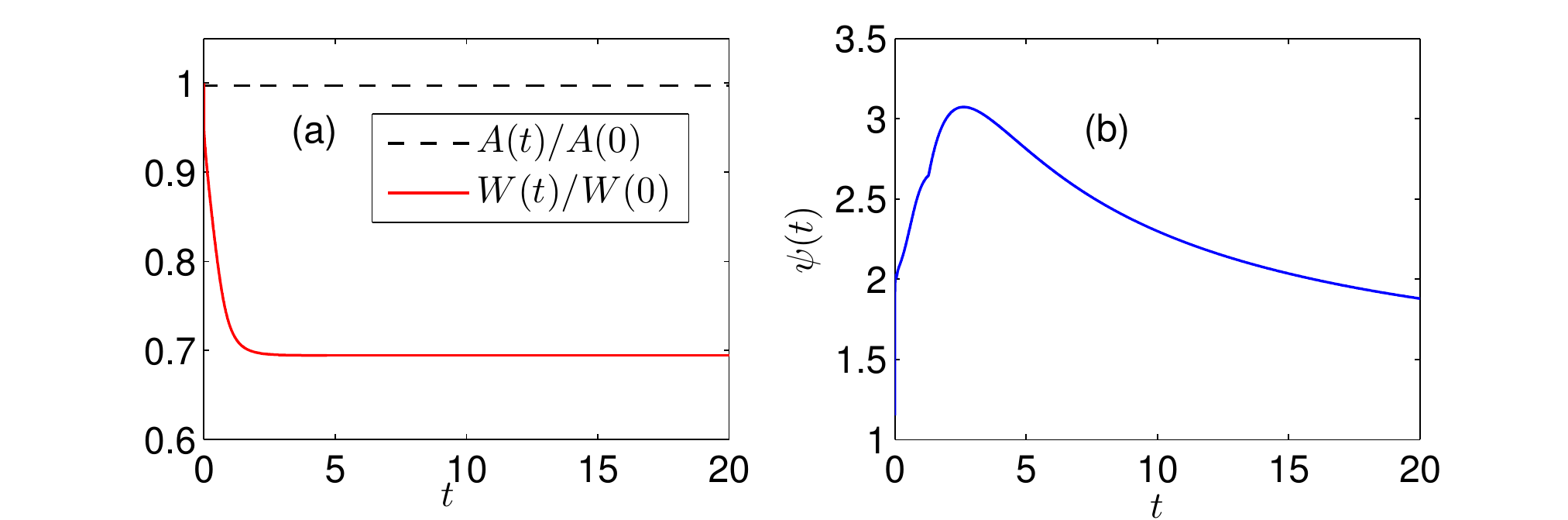}
\caption{\label{fig:wkmbetaME} (a) The temporal evolution of the normalized total free energy and the normalized total  area/mass; (b) the temporal evolution of the mesh distribution function $\psi(t)$. The computational parameters are chosen as the same as Fig.~\ref{fig:wkmbeta1}(c).}
\end{figure}

Fig.~\ref{fig:wkmbetaME}(a) shows the temporal evolution of the normalized free energy $W(t)/W(0)$ and the normalized area/mass $A(t)/A(0)$ defined in the previous section. As clearly shown in Fig.~\ref{fig:wkmbetaME}(a), the horizontal black dash line implies that our PFEM has a very good property which ensures that the total area/mass of the thin film conserves, and the monotonically decreasing red solid line implies the energy dissipation property during the whole evolution process. We also investigate the mesh quality by defining the mesh distribution function $\psi(t)$ as
\begin{equation}
\psi(t_m) = \frac{\max_{1\le j\le N}|\vec h_j^m|}{\min_{1\le j\le N}|\vec h_j^m|},\quad\text{where}\;\Gamma^m=\cup_{j=1}^N\vec h_j^m.
\end{equation}

Fig.~\ref{fig:wkmbetaME}(b) shows the temporal evolution of the mesh distribution function $\psi(t)$. As shown in the figure, we can clearly observe that the distribution function first quickly increases from $1$ to about $3$ and then gradually decreases to a value around 2, and its value is always not big during the evolution. We find that the mesh quality is always preserved well during the simulation when the surface energy anisotropy is not very strong (i.e., $\beta$ is not very big). Some theoretic analysis for the mesh-distribution property in the isotropic surface energy case can be found in the reference~\cite{Barrett07Iso}, and an intuitive explanation is because Eq.~\eqref{eqn:weakfull1} allows the tangential velocity of mesh points which does not change the shape of the interface curve, while this tangential velocity tends to distribute the mesh points uniformly according to the arc length due to Eq.~\eqref{eqn:weakfull2}.

\begin{figure}[htp!]
\centering
\includegraphics[width=0.90\textwidth]{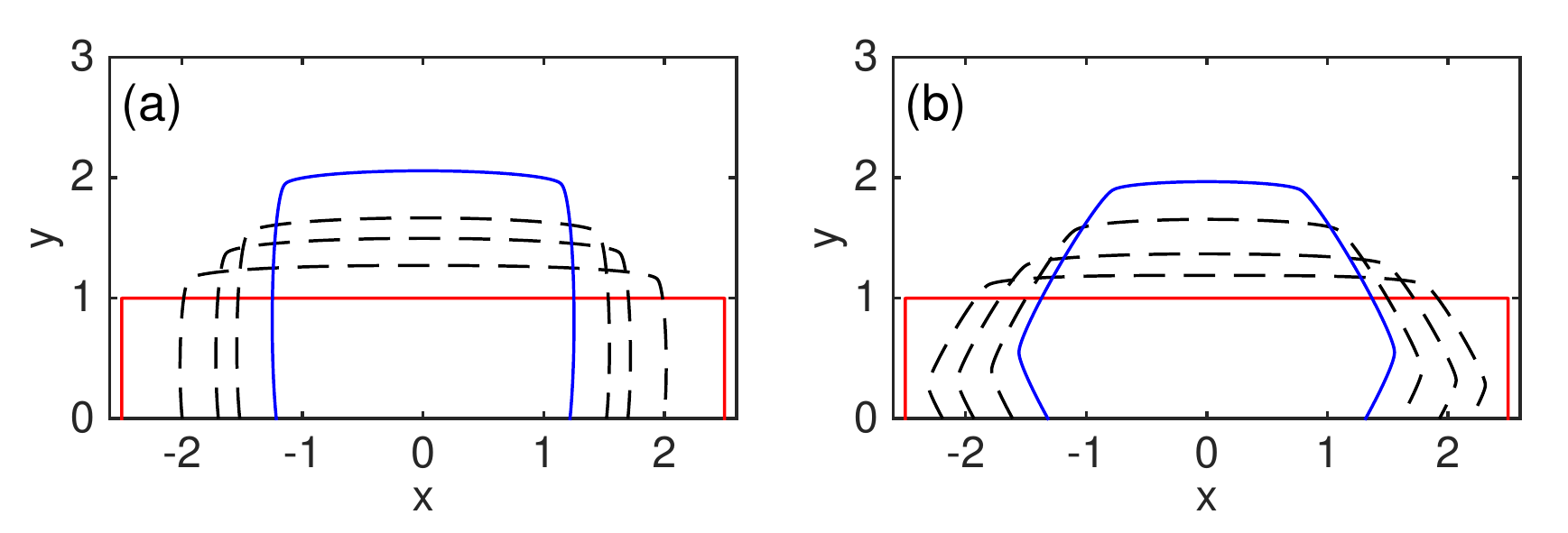}
\caption{Several steps in the evolution of small, initially rectangular islands (shown in red) towards their
equilibrium shapes (shown in blue) where the anisotropy is given by Eq.~\eqref{eqn:cuspsenergy},
where the computational parameters are chosen as $\sigma=\cos\frac{3\pi}{4}$, and the anisotropic parameters are: (a) $K=2, \phi_1=0,\phi_2=\frac{\pi}{2},\delta_1=\delta_2=0.1$, (b) $K=3,\phi_1=0,\phi_2=\frac{\pi}{3},\phi_3=\frac{2\pi}{3},\delta_1=\delta_2=\delta_3=0.1$, and the number of grid points is chosen as $N=800$, the time step is uniformly chosen as $\tau=10^{-4}$.}
\label{fig:wkcusps}
\end{figure}

The next type of surface energy density we will focus on is of the Riemannian metric form defined in Eq.~\eqref{eqn:riemann}, i.e.,
\begin{equation}
\gamma(\vec n) = \sum_{k=1}^K\sqrt{ G_k \vec n\cdot\vec n},
\quad G_k=R(-\phi_k)D(\delta_k)R(\phi_k), \quad k=1,\cdots,K,
\label{eqn:cuspsenergy}
\end{equation}
where the matrices $D$ and $R$ are defined as
\begin{equation}
D(\delta)=
\begin{pmatrix}
1 & 0 \\ 0 &  \delta^2
\end{pmatrix},\qquad
R(\phi)=
\begin{pmatrix}
\cos\phi& \sin\phi\\-\sin\phi &\cos\phi
\end{pmatrix}.
\end{equation}
Here, the matrix $D$ is positive definite, and the regularization parameter $\delta$ can be viewed as a kind of smooth regularization for this type of surface energy anisotropy (i.e., used for smoothing sharp corners which will appear in equilibrium shapes). When $\delta$ decrease from a small positive number to zero, its corresponding equilibrium shape will exhibit sharper and sharper corners.

Under the above type of surface energy, we perform numerical simulations for investigating the kinetic evolution of solid-state dewetting. Fig.~\ref{fig:wkcusps} shows the kinetic evolution of a small initial rectangular island (in red) towards its equilibrium shape (in blue), where the material constant is chosen as $\sigma=\cos\frac{3\pi}{4}$, and the parameters that control the surface energy anisotropy are chosen as: (a) $K=2,\phi_1=0,\phi_2=\frac{\pi}{2},\delta_1=\delta_2=0.1$, and (b) $K=3,\phi_1=0,\phi_2=\frac{\pi}{3},\phi_3=\frac{2\pi}{3},\delta_1=\delta_2=\delta_3=0.1$.
As clearly shown in Fig.~\ref{fig:wkcusps}, it can be observed that the evolution shape for this type of anisotropy seems to be more ``faceting'' than the smooth $k$-fold anisotropy, and the equilibrium shape for Fig.~\ref{fig:wkcusps}(a) is a truncation of a square while it is a truncation of a hexagon for Fig.~\ref{fig:wkcusps}(b), which implies that the parameter $K$ plays the same role in determining the symmetry as it does in the smooth $k$-fold anisotropy. The small regularization parameter $\delta$ is here used to smoothen sharp corners which connect with two different facets.
\begin{figure}[!htp]
\centering
\includegraphics[width=0.8\textwidth]{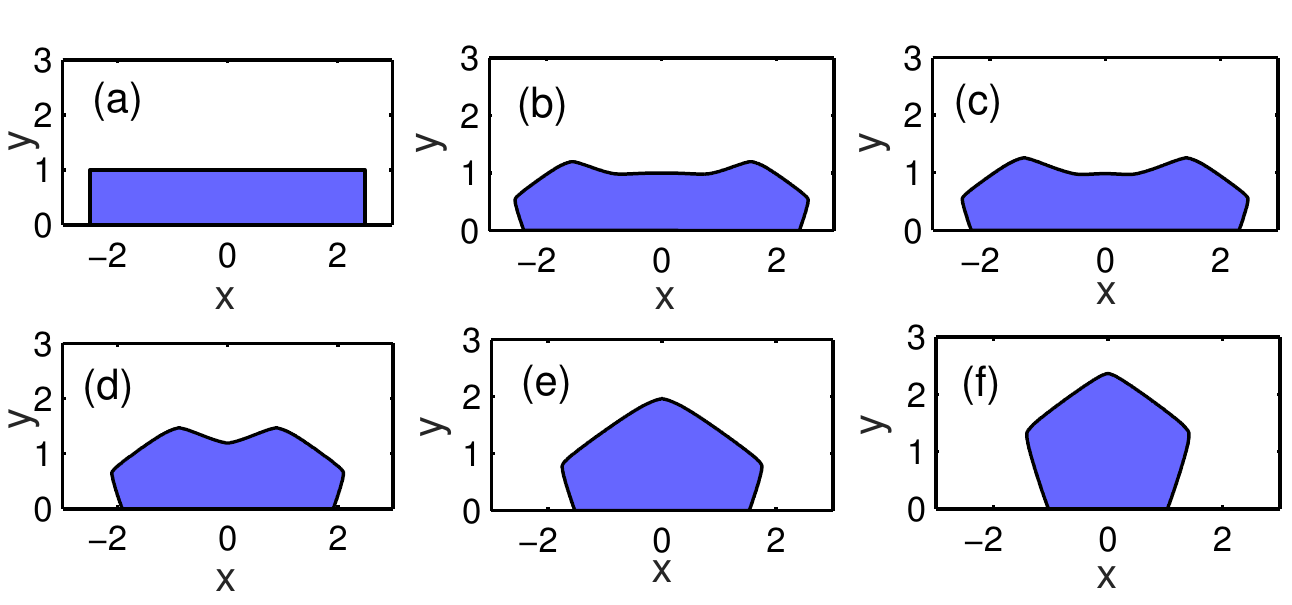}
\caption{\label{fig:absolute} Several snapshots in the evolution of a small, initially rectangular island film towards its equilibrium shape at different times:(a) $t=0$; (b) $t=0.05$; (c) $t=0.1$; (d) $t=0.5$; (e) $t=1.0$; (f) $t=7.7$, where the surface energy density is chosen as $\gamma(\theta)=1+0.19\,|\cos(\frac{5}{2}\theta)\,|$ with a small regularization parameter $\delta=0.1$, and the material constant $\sigma=\cos\frac{3\pi}{4}$, the stabilized parameter in the stabilized PFEM scheme \eqref{eqn:weakfull5} -\eqref{eqn:weakfull6} is chosen as $\lambda=20$, and the number of grid points is chosen as $N=800$, the time step is uniformly chosen as $\tau=10^{-4}$.}
\end{figure}

The last type of surface energy anisotropy we consider in this section is defined as~\cite{Barrett11},
\begin{equation}
\gamma(\theta) = 1 + \beta\,|\cos \frac{k\theta}{2} \,|.
\end{equation}
and this type of surface energy is not smooth at the points where the function $\cos\frac{k\theta}{2}$ changes its sign. A smooth regularization technique can be done as follows because the proposed PFEM needs $\gamma(\theta) \in C^1[-\pi,\pi]$,
\begin{equation}
\gamma(\theta) = 1 + \beta \sqrt{\delta^2 + \cos^2 \frac{k\theta}{2}},
\end{equation}
where $\beta$ controls the degree of anisotropy. During our practical simulations, we find that in this case when $\beta$ becomes very large (but still in the weakly anisotropic, i.e., $\gamma(\theta)+\gamma\,''(\theta)>0$), the PFEM scheme \eqref{eqn:weakfull1}-\eqref{eqn:weakfull3} does not work better than the stabilized PFEM scheme \eqref{eqn:weakfull5}-\eqref{eqn:weakfull6} in the sense that we use the same number of grid points. So here, we use the stabilized PFEM scheme \eqref{eqn:weakfull5}-\eqref{eqn:weakfull6} for simulating solid-state dewetting with this special type of anisotropy. As shown in Fig.~\ref{fig:absolute}, the interface curve evolves from a small rectangular island to
its equilibrium shape (a regular $k$-polygon truncated by a flat substrate). Here, the computational parameter about this type of anisotropy is chosen as $\beta=0.19$, $k=5$, $\delta=0.1$ and the stabilized parameter is chosen as $\lambda=20$.

\subsection{Large islands}

As widely discussed in the papers~\cite{Dornel06,Jiang12,Wang15,Bao17}, when the aspect ratios of thin island films are larger than a critical value, the large islands will pinch-off to form small separated islands.

\begin{figure}[!htp]
\centering
\includegraphics[width=0.9\textwidth]{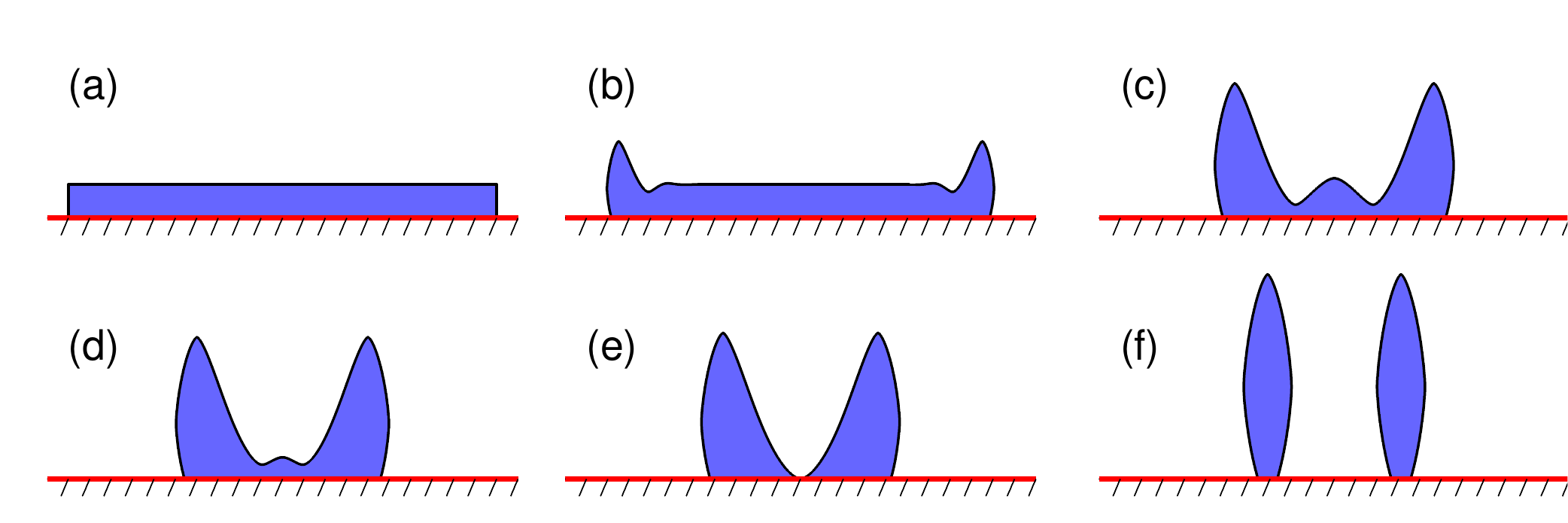}
\caption{Several snapshots in the evolution of a long, thin island film (aspect ratio of $60$) with $k$-fold
smooth anisotropic surface energy ($k=4,\beta=0.06$) and the material parameter $\sigma=\cos(5\pi/6)$ at different times: (a)~$t=0$; (b)~$t=10$; (c)~$t=240$; (d)~$t=320$; (e)~$t=371$; (f)~$t=711$. Note the difference in vertical and horizontal scales.}
\label{fig:weaklarge}
\end{figure}

In order to obtain a qualitative comparison with other numerical methods, we choose the same computational parameters as in the paper~\cite{Wang15,Bao17}. The numerical computation is set up as follows: the initial thin film is chosen as a very large thin island with length $L=60$ and height $h=1$. The anisotropy is given as $4$-fold anisotropic surface energy density with $\beta=0.06$ and the material constant is given by $\sigma=\cos5\pi/6$.

\begin{figure}[!htp]
\centering
\includegraphics[width=0.8\textwidth]{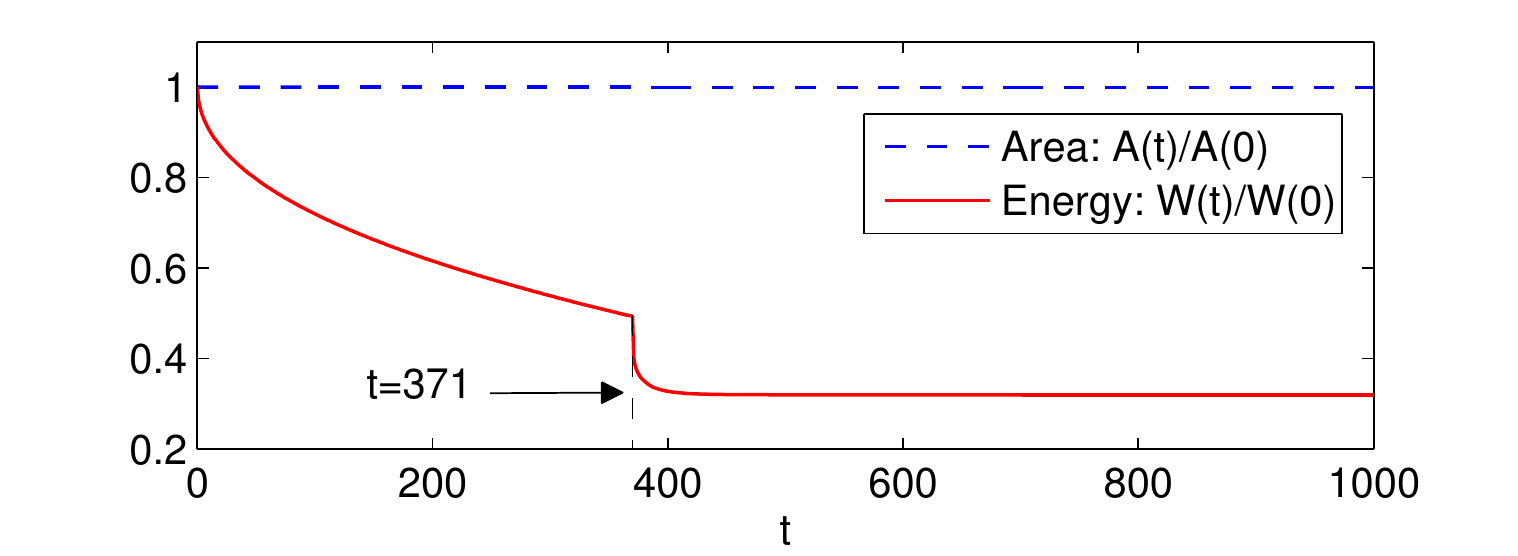}
\caption{The corresponding temporal evolution in Fig.~\ref{fig:weaklarge} for the normalized total
free energy and the normalized area (mass).}
\label{fig:weaklongenergy}
\end{figure}

Fig.~\ref{fig:weaklarge} depicts the temporal geometric evolution of this initially rectangular island during the solid-state dewetting.  As can be seen in the figure,
solid-state dewetting very quickly leads to the formation of ridges at the island edges followed by two valleys.
As time evolves, the ridges and valleys become increasingly exaggerated, then the two valleys merge near the island center. At the time $t=371$, the valley at the center of the island hits the substrate, leading to a pinch-off
event that separates the initial island into a pair of islands. Finally, the two separated islands continue to evolve until they reach their equilibrium shapes. The corresponding evolution of the normalized total free energy and normalized
total area (mass) are shown in Fig.~\ref{fig:weaklongenergy}. An interesting phenomenon here is that the total energy
undergoes a sharp drop at $t=371$, the moment when the pinch-off event occurs.

 The pinch-off time $t=371$ we obtained by using our PFEM for this particular example is very close to the result $t=374$ by using marker-particle methods in~\cite{Wang15}. But under the same computational resource, the computational time by using PFEM for this example is about two hours, while it is about two weeks by using marker-particle methods~\cite{Wang15}. Besides, the obtained pinch-off time is exactly the same as the result by using another PFEM method recently proposed in~\cite{Bao17} which does not use $\boldsymbol{\xi}$-vector and has more unknown variables in its variational form, and it validates the accuracy of the new PFEM from one side. Note here that once the interface curve hits the substrate somewhere in the simulation, it means that a pinch-off event has happened and a new contact point is generated, then after the pinch-off, we compute each part of the pinch-off curve separately.

\begin{figure}[!htp]
\centering
\includegraphics[width=0.9\textwidth]{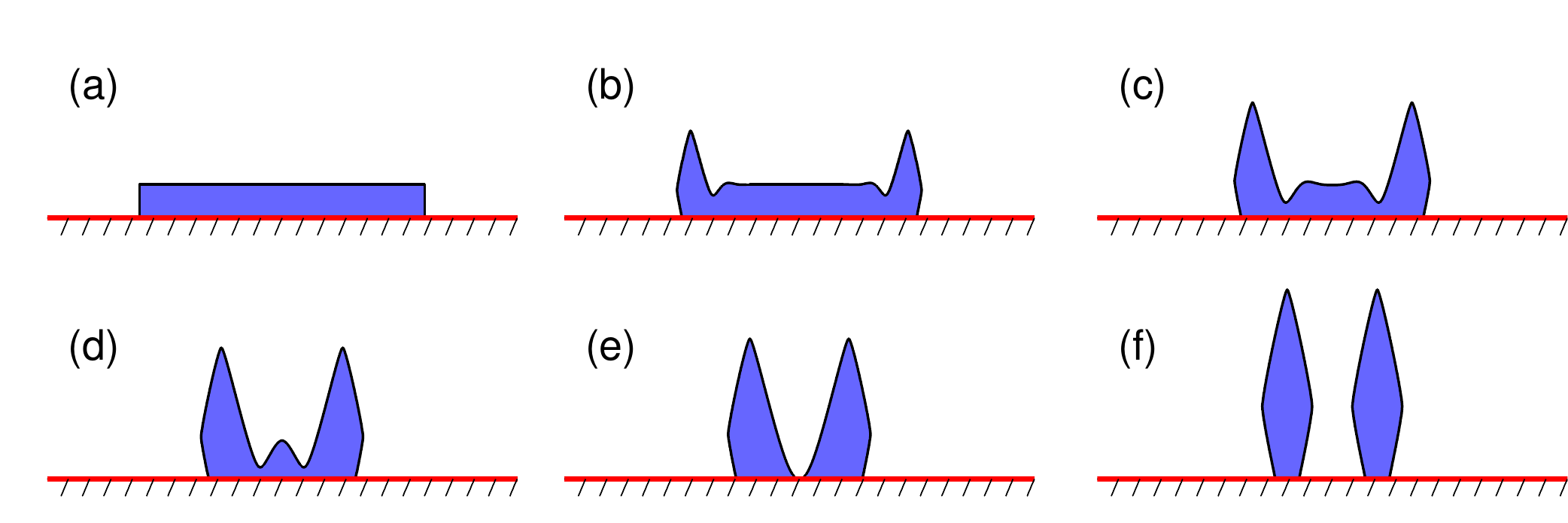}
\caption{Several snapshots in the evolution of a long, thin island film (aspect ratio of $40$) for surface energy defined in Eq.~\eqref{eqn:cuspsenergy}, with $K=2,\phi_1=\pi/4,\phi_2=3\pi/4,\delta_1=\delta_2=0.1$. And the material parameter $\sigma=\cos(5\pi/6)$:(a)~$t=0$; (b)~$t=10$; (c)~$t=50$; (d)~$t=100$; (e)~$t=140$; (f)~$t=400$. Note the difference in vertical and horizontal scales.}
\label{fig:weaklarge1}
\end{figure}
The pinch-off event is not limited to the smooth $k$-fold anisotropy, and it can also be observed for other types of surface energy, e.g., Eq.~\eqref{eqn:cuspsenergy}. Fig.~\ref{fig:weaklarge1} shows the geometric evolution of a large, initially rectangular island with aspect ratio of $40$. The material constant $\sigma$ is chosen as the same as Fig.~\ref{fig:weaklarge}. The parameters which control the surface energy anisotropy are chosen as $K=2,\phi_1=\pi/4,\phi_2=3\pi/4,\delta_1=\delta_2=0.1$. As shown in the figure, we can clearly observe that the thin film quickly forms valleys and ridges at its edges and then the pinch-off event happens at about the time $t=140$. Subsequently, the thin film breaks up into two small thin island pieces which finally evolve into their corresponding equilibrium shapes. This evolution process shares a similar geometric evolution with the smooth $k$-fold anisotropy.

\subsection{Semi-infinite films}

\begin{figure}[!htp]
\centering
\includegraphics[width=0.9\textwidth]{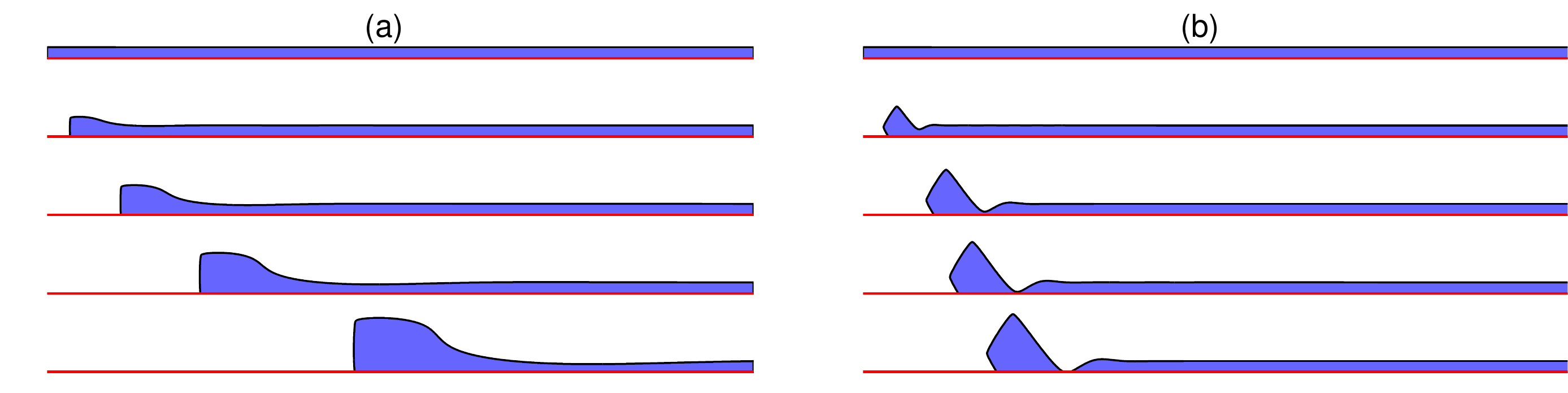}
\caption{\label{fig:cusp1} The temporal evolution of a semi-infinite step film under the form of surface energy anisotropy defined in Eq.~\eqref{eqn:cuspsenergy} with different parameters, (a) Case A:~$(\phi_1,\phi_2)=(0,\frac{\pi}{2}),\delta_1=\delta_2=0.1$, (b) Case B: ~$(\phi_1,\phi_2)=(\frac{\pi}{4},\frac{3\pi}{4}),\delta_1=\delta_2=0.1$.
The material constant is chosen as $\sigma=\cos \frac{5\pi}{6}$.}
\end{figure}

For the retraction of a semi-infinite step film, a lot of earlier studies have shown that the retraction distance $R(t)$ of a semi-infinite step film as a function of time satisfies a power-law relation \cite{Srolovitz86,Kim13,Wong00,Wang15,Zucker16power}, i.e., $R(t)\sim t^{\alpha}$. But most of the above studies focused on the isotropic surface energy case~\cite{Srolovitz86,Wong00,Jiang12} or some specific forms of surface energy anisotropy~\cite{Kim13,Wang15,Zucker16power}. For any form of surface energy anisotropy, does the power law exponent $\alpha$ depend on the type of surface energy anisotropy? this is still a question. Here, we want to investigate this power-law relation by performing ample numerical simulations on semi-infinite thin step films with the surface energy anisotropy defined in Eq.~\eqref{eqn:cuspsenergy}. As illustrated in Fig.~\ref{fig:cusp1}, we simulate the retraction evolution process of a semi-infinite step film in two different cases. In both cases, the material constant is chosen as $\sigma=\cos\frac{5\pi}{6}$, while the anisotropies are chosen as two different parameters: for Case A, $\gamma(\vec n)$ is chosen as the form of Eq.~\eqref{eqn:cuspsenergy} with the parameters $(\phi_1,\phi_2)=(0,\frac{\pi}{2}),\delta_1=\delta_2=0.1$; for Case B, $\gamma(\vec n)$ is chosen as the form of Eq.~\eqref{eqn:cuspsenergy} with the parameters $(\phi_1,\phi_2)=(\frac{\pi}{4},\frac{3\pi}{4}),\delta_1=\delta_2=0.1$. The surface energy density in Case B can be viewed as the rotation of $\frac{\pi}{4}$ for Case A.


\begin{figure}[!htp]
\centering
\includegraphics[width=0.9\textwidth]{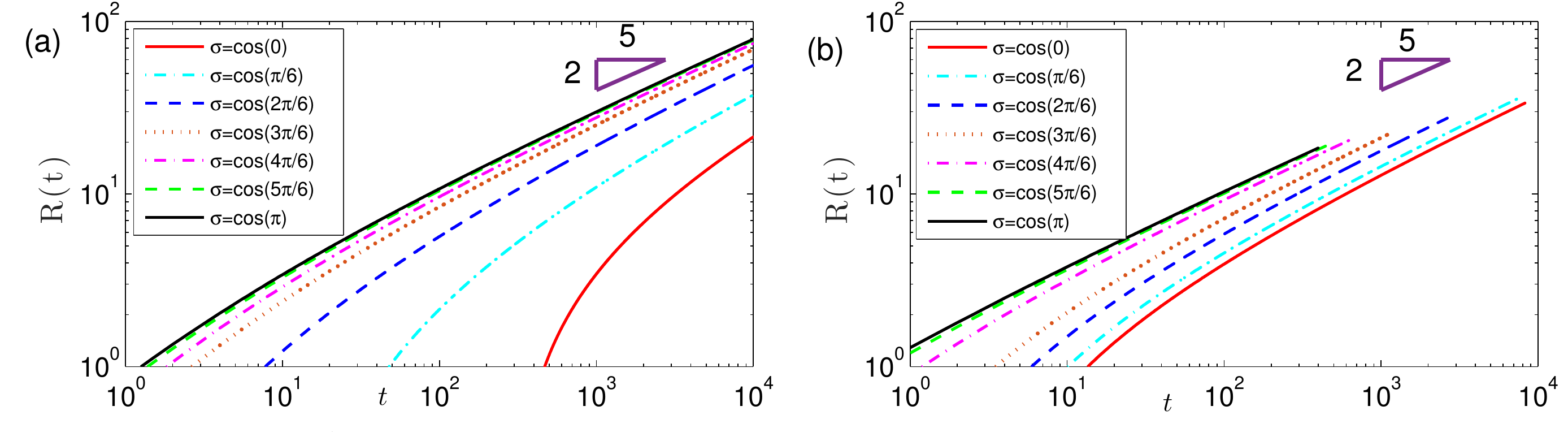}
\caption{The retraction distance versus time $t$ of a semi-infinite step film with the form of surface energy anisotropy defined in Eq.~\eqref{eqn:cuspsenergy} under different material constants $\sigma$, (a):~$(\phi_1,\phi_2)=(0,\frac{\pi}{2}),\delta_1=\delta_2=0.1$, (b): ~$(\phi_1,\phi_2)=(\frac{\pi}{4},\frac{3\pi}{4}),\delta_1=\delta_2=0.1$.}
\label{fig:semirectration1}
\end{figure}

As is clearly shown in Fig.~\ref{fig:cusp1}(a), a ridge quickly forms at the retracting edge but no valley appears behind
the ridge, and the whole semi-infinite step film gradually moves towards the right direction along the substrate. As time evolves, the ridge becomes higher and higher, but because no valley forms, the pinch-off will never happen.
This can be explained as because the surface energy density attains its minimum at the orientations $\vec n =(0,1), (1,0),(-1,0)$, and the step film quickly evolves to its ``quasi-equilibrium'' shape, i.e., preferring its minimum energy
orientations, and the oblique direction is not the most favorable energy directions, so valleys never form for this kind of surface energy. On the other hand, if we use another kind of surface energy (e.g. taking a $45$ degrees rotation to
the minimal energy orientations), as shown in Fig.~\ref{fig:cusp1}(b), a valley will quickly form behind the ridge, and
as time evolves, the valley sinks with time and eventually touches with the substrate, then the pinch-off will happen.
It should be noted that, for isotropic surface energy, a lot of research works have shown that valleys always form ahead of retracting ridges and can eventually lead to the pinch-off phenomena~\cite{Wong00,Dornel06,Jiang12}. However, for anisotropic surface energy, the situation is totally different, and valleys may appear or be absent at retracting edges according to different surface energies (shown in Fig.~\ref{fig:cusp1}). This observation is consistent with experimental studies for single crystal thin films, and the formation of valleys ahead of the ridges can be observed or not observed in real experiments with different set-ups~\cite{Kim13,Zucker16}.

Figs.~\ref{fig:semirectration1}(a)-(b) depict the log-log plots of the retraction distance $R(t)$ versus time $t$ under different material constants $\sigma$, and the surface energy is chosen as the form of Eq.~\eqref{eqn:cuspsenergy} with different controlled parameters: (a)~$(\phi_1,\phi_2)=(0,\frac{\pi}{2}),\delta_1=\delta_2=0.1$; (b) $(\phi_1,\phi_2)=(\frac{\pi}{4},\frac{3\pi}{4}),\delta_1=\delta_2=0.1$. The numerical results clearly show that the retraction distance $R(t)$ can be well described by a power-law relation, and the power-law exponent $\alpha\approx 2/5$ is insensitive to the material constant and surface energy anisotropy.
\section{Extensions to strongly anisotropic case}

Sharp corners would appear in the equilibrium shape when the surface stiffness $\rm{H}_\gamma(\vec n)\boldsymbol{\tau}\cdot\boldsymbol{\tau}=\gamma(\theta)+\gamma^{\prime\prime}(\theta)<0$ for some orientations $\theta$, i.e., the strongly anisotropic case. In this case, the sharp-interface governing equations \eqref{eqn:weak1}-\eqref{eqn:weak2} becomes ill-posed.
These governing equations can be regularized by adding regularization terms such that the regularized sharp-interface model is well-posed. In this section, we regularize the total interfacial energy $W(\Gamma)$ defined in Eq.~\eqref{eqn:surfaceenergy} by adding the well-known Willmore energy~\cite{Gurtin02,li09}, which is defined as
\begin{equation}
 W_{\rm {wm}}=\frac{\varepsilon^2}{2}\int_{\Gamma}\kappa^2\;ds,
 \end{equation}
where $\kappa$ is the curvature of the interface curve and $0<\varepsilon \ll 1$ is a small regularization parameter. By adding this regularization term, the total interfacial energy can be written as follows
\begin{equation}\label{eqn:strongenergy}
W_{\rm {reg}}^\varepsilon(\Gamma)= W(\Gamma)+W_{\rm {wm}}(\Gamma)=\int_{\Gamma}(\gamma(\vec n)+\frac{\varepsilon^2}{2}\kappa^2)\;ds  - (\gamma_{_{\subVS}}-\gamma_{_{\subFS}})(x_c^r-x_c^l).
\end{equation}

\subsection{The (regularized) energy and first variation}
In order to perform the first variation of the total surface energy defined in Eq.~\eqref{eqn:strongenergy}, we need the following lemma.

\begin{lem} \label{le:stopencurve}Assume that $\Gamma=\vec X(s)\in C^{4}([0,L])\times C^4([0,L])$ is an open curve
with its two endpoints locating at $s=0$ and $s=L$, where $s:=s(\rho)$ represents the arc length of the curve.
Consider a perturbation for $\Gamma$ with $\vec V(\rho,\epsilon)$ representing the smooth variational vector field defined in Eq.~\eqref{eqn:Vectorfield}. Define the shape functional $F(\Gamma)=\int_{\Gamma}\kappa^2 \;ds$, then we have
\begin{equation}
\delta F(\Gamma;\vec V) = -\int_{\Gamma}(\kappa^3 + 2\partial_{ss}\kappa)\,(\vec V_0\cdot \vec n\;)ds-\Big(2\kappa\,(\vec n\cdot\partial_s\vec V_0) - 2\partial_s\kappa\,(\vec V_0 \cdot \vec n) + \kappa^2\,(\vec V_0 \cdot \boldsymbol{\tau} )\Big)\Big|_{s=0}^{s=L},
\end{equation}
where the deformation velocity is denoted as $\vec V_0=\vec V(\rho, 0)$, and $\vec V_0 \cdot \vec n$ represents the deformation velocity along the outer normal direction of the interface.
\begin{proof}
If we define $\theta^\epsilon:=\theta(\rho,\epsilon)=\arctan\frac{\partial_\rho y(\rho,\epsilon)}{\partial_\rho x(\rho,\epsilon)}$, then $\kappa^\epsilon:=\kappa(\rho,\epsilon)$ with respect to the perturbed curve can be expressed as
\begin{equation}
\kappa^\epsilon = -\partial_{s^\epsilon}\theta^\epsilon= -\frac{\partial_\rho\theta^\epsilon}{|\partial_\rho\vec X(\rho,\epsilon)|}.
\label{eqn:kappatheta}
\end{equation}
From the parameterization defined in Eq.~\eqref{eqn:parameterization}, $F(\Gamma^\epsilon)$ is given by
\begin{equation}
F(\Gamma^\epsilon) = \int_0^1(\kappa^\epsilon)^2|\partial_\rho\vec X(\rho,\epsilon)|\;d\rho.
\end{equation}
By taking derivative of $F(\Gamma^\epsilon)$ with respect to $\epsilon$ and noting the independence between $\epsilon$ and $\rho$,  based on Eq.~\eqref{eqn:kappatheta}, we have
\begin{eqnarray}
\frac{d}{d\epsilon}F(\Gamma^\epsilon)&=&\int_0^1\,2\kappa^\epsilon\partial_\epsilon\kappa^\epsilon|\partial_\rho\vec X(\rho,\epsilon)|\;d\rho+\int_0^1(\kappa^\epsilon)^2\partial_\epsilon|\partial_\rho\vec X(\rho,\epsilon)|\;d\rho\nonumber\\
&=&-2\int_0^1\kappa^\epsilon\partial_\epsilon\partial_\rho\theta^\epsilon\;d\rho -2 \int_0^1\,\kappa^\epsilon\partial_\rho\theta^\epsilon\partial_\epsilon\frac{1}{|\partial_\rho\vec X(\rho,\epsilon)|}|\partial_\rho\vec X(\rho,\epsilon)|\;d\rho + \int_0^1(\kappa^\epsilon)^2\partial_\epsilon|\partial_\rho\vec X(\rho,\epsilon)|\;d\rho\nonumber\\
&=& I+II+III.
\end{eqnarray}
Note that performing the Taylor expansion of $\theta^\epsilon$ and $\frac{1}{|\partial_\rho\vec X(\rho,\epsilon)|}$ with respect to the small perturbation parameter at $\epsilon=0$, then we obtain
\begin{subequations}
\begin{align}
&\theta^\epsilon = \theta - \frac{\partial_\rho\vec X^\perp\cdot\partial_\rho\vec V_0}{|\partial_\rho\vec X|^2}\epsilon + O(\epsilon^2)=\theta + \theta'\epsilon + O(\epsilon^2).\label{eqn:prim}\\
&\frac{1}{|\partial_\rho\vec X(\rho,\epsilon)|} = \frac{1}{|\partial_\rho\vec X|} - \frac{\partial_\rho\vec X\cdot\partial_\rho\vec V_0}{|\partial_\rho\vec X|^3}\epsilon + O(\epsilon^2).
\end{align}
\end{subequations}
Using above expansions and Eq.~\eqref{eqn:perb1}, and for the above three parts, by taking the values at $\epsilon=0$ and integration by parts, we have
\begin{eqnarray}
I\Big|_{\epsilon=0}&=&\int_0^12\kappa(-\partial_\rho\theta^\prime)\;d\rho
=-\Bigl(2\kappa\theta'\Bigr)\Big|_{\rho=0}^{\rho=1}+\int_0^12\partial_\rho\kappa\;\theta^\prime\;d\rho\nonumber\\
&=&-\Bigl(2\kappa\theta'\Bigr)\Big|_{\rho=0}^{\rho=1}-\int_0^12\partial_\rho\kappa\Bigl(\frac{\partial_\rho\vec X^{\perp}\cdot\partial_\rho\vec V_0}{|\partial_\rho\vec X|^2}\Bigr)\;d\rho\nonumber\\
&=&-\Bigl(2\kappa\theta'\Bigr)\Big|_{\rho=0}^{\rho=1}-\Bigl(\frac{2\partial_\rho\kappa\partial_\rho\vec X^{\perp}\cdot\vec V_0}{|\partial_\rho\vec X|^2}\Bigr)\Big|_{\rho=0}^{\rho=1}+\int_{0}^1\partial_\rho\Bigl(\frac{2\partial_\rho\kappa\partial_\rho\vec X^{\perp}}{|\partial_\rho\vec X|^2}\Bigr)\cdot\vec V_0\;d\rho\nonumber\\
&=&-\Bigl(2\kappa\theta'\Bigr)\Big|_{\rho=0}^{\rho=1}-\Bigl(\frac{2\partial_\rho\kappa\partial_\rho\vec X^{\perp}\cdot\vec V_0}{|\partial_\rho\vec X|^2}\Bigr)\Big|_{\rho=0}^{\rho=1}+\int_{\Gamma}(2\partial_{ss}\kappa\,\partial_s\vec X^{\perp}+2\partial_s\kappa\,\partial_{ss}\vec X^{\perp})\cdot\vec V_0\;ds\nonumber\\
&=&\Bigl(-2\kappa\theta' + 2\partial_s\kappa\,\vec n\cdot\vec V_0\Bigr)\Big|_{s=0}^{s=L}-2\int_{\Gamma}(\partial_{ss}\kappa\,\vec n + \partial_s\kappa\,\kappa\,\boldsymbol{\tau})\cdot\vec V_0\;ds.
\end{eqnarray}
\begin{eqnarray}
II\Big|_{\epsilon=0}&=&\int_0^12\kappa\partial_\rho\theta\frac{\partial_\rho\vec X\cdot\partial_\rho\vec V_0}{|\partial_\rho\vec X|^2}\;d\rho=2\Bigl(\kappa\partial_\rho\theta\frac{\partial_\rho\vec X\cdot\vec V_0}{|\partial_\rho\vec X|^2}\Bigr)\Big|_{\rho=0}^{\rho=1}-\int_0^12\partial_\rho\Bigl(\kappa\partial_\rho\theta\frac{\partial_\rho\vec X}{|\partial_\rho\vec X|^2}\Bigr)\vec V_0\;d\rho\nonumber\\
&=&2\Bigl(\kappa\partial_\rho\theta\frac{\partial_\rho\vec X\cdot\vec V_0}{|\partial_\rho\vec X|^2}\Bigr)\Big|_{\rho=0}^{\rho=1}+\int_{\Gamma}2\partial_s(\kappa^2\partial_s\vec X)\cdot\vec V_0\;ds\nonumber\\
&=&2\Bigl(\kappa\partial_\rho\theta\frac{\partial_\rho\vec X\cdot\vec V_0}{|\partial_\rho\vec X|^2}\Bigr)\Big|_{\rho=0}^{\rho=1}+\int_{\Gamma}(4\kappa\partial_s\kappa\partial_s\vec X+2\kappa^2\partial_{ss}\vec X)\cdot\vec V_0\;ds\nonumber\\
&=&-2\Bigr(\kappa^2\boldsymbol{\tau}\cdot\vec V_0\Bigr)\Big|_{s=0}^{s=L} + 2\int_{\Gamma}(2\kappa\,\partial_s\kappa\,\boldsymbol{\tau} - \kappa^3\vec n)\cdot\vec V_0\;ds.
\end{eqnarray}
\begin{eqnarray}
III\Big|_{\epsilon=0}&=&\int_0^1\kappa^2\frac{\partial_\rho\vec X\cdot\partial_\rho\vec V_0}{|\partial_\rho\vec X|}\;d\rho=\Bigl(\kappa^2\frac{\partial_\rho\vec X\cdot\vec V_0}{|\partial_\rho\vec X|}\Bigr)_{\rho=0}^{\rho=1}-\int_0^1\partial_\rho\Bigl(\kappa^2\frac{\partial_\rho\vec X}{|\partial_\rho\vec X|}\Bigr)\vec V_0\;d\rho\nonumber\\
&=&\Bigl(\kappa^2\frac{\partial_\rho\vec X\cdot\vec V_0}{|\partial_\rho\vec X|}\Bigr)\Big|_{\rho=0}^{\rho=1}-\int_{\Gamma}\partial_s\Bigl(\kappa^2\partial_s\vec X\Bigr)\cdot\vec V_0\;ds\nonumber\\
&=&\Bigl(\kappa^2\frac{\partial_\rho\vec X\cdot\vec V_0}{|\partial_\rho\vec X|}\Bigr)\Big|_{\rho=0}^{\rho=1}-\int_{\Gamma}(2\kappa\partial_s\kappa\partial_s\vec X+\kappa^2\partial_{ss}\vec X)\cdot\vec V_0\;ds\nonumber\\
&=&\Bigl(\kappa^2 \boldsymbol{\tau}\cdot\vec V_0\Bigr)\Big|_{s=0}^{s=L}-\int_{\Gamma}(2\kappa\,\partial_s\kappa\,\boldsymbol{\tau} - \kappa^3\vec n)\cdot\vec V_0\;ds.
\end{eqnarray}
Making the summation for the above three terms, we can obtain
\begin{equation}
(I+II+III)\Big|_{\epsilon=0}
=-\int_{\Gamma}(\kappa^3+2\partial_{ss}\kappa)\,(\vec n\cdot\vec V_0)\;ds-\Bigl(2\kappa\theta'\Bigr)\Big|_{s=0}^{s=L}+\Bigl((2\partial_s\kappa\vec n - \kappa^2\boldsymbol{\tau})\cdot\vec V_0\Bigr)\Big|_{s=0}^{s=L}.
\end{equation}
Note that the notation $\theta'$ is defined in Eq.~\eqref{eqn:prim}, i.e.,
\begin{equation}
\theta'=-\frac{\partial_\rho\vec X^{\perp}\cdot\partial_\rho\vec V_0}{|\partial_\rho\vec X|^2}=\vec n\cdot\partial_s\vec V_0,
\end{equation}
thus we have
\begin{equation}
\delta F(\Gamma;\vec V) = -\int_{\Gamma}(\kappa^3 + 2\partial_{ss}\kappa)\,(\vec n\cdot\vec V_0\;)ds-\Bigl(2\kappa\,(\vec n\cdot\partial_s\vec V_0) - 2\partial_s\kappa\,(\vec n\cdot\vec V_0) + \kappa^2\,(\boldsymbol{\tau}\cdot\vec V_0)\Bigr)\Big|_{s=0}^{s=L}.
\end{equation}
\end{proof}
\end{lem}
\subsection{The (regularized) sharp-interface model}

Now, from the above Lemma.~\ref{le:stopencurve}, the first variation of the regularized Willmore energy term can be written as follows
\begin{equation}
\delta W_{{\rm {wm}}}(\Gamma;\vec V)
=-\varepsilon^2\int_{\Gamma}(\frac{\kappa^3}{2}+\partial_{ss}\kappa)\,(\vec n\cdot\vec V_0)\;ds-\,\varepsilon^2\Bigl(\kappa\,(\vec n\cdot\partial_s\vec V_0) - \partial_s\kappa\,(\vec n\cdot\vec V_0) + \frac{\kappa^2}{2}\,(\boldsymbol{\tau}\cdot\vec V_0)\Bigr)\Big|_{s=0}^{s=L}.
\end{equation}
The boundary term $-\Bigl(\kappa(\vec n\cdot\partial_s\vec V_0)\Bigr)\Big|_{s=0}^{s=L}$ tells us to impose the zero curvature boundary condition into the model to ensure the energy dissipation for arbitrary perturbation.
Because the contact points must move along the substrate, the perturbation velocity field at the contact points must satisfy  the relation: $\vec V_0(s=0)\sslash \vec e_1$ and $\vec V_0(s=L)\sslash \vec e_1$, where ${\vec e_1}=(1,0)$ represents the unit vector along the $x$-coordinate (or the substrate line). Making use of the boundary condition $\kappa=0$ at the two contact points and by combining Eq.~\eqref{eqn:DlessEgva} with the above equation, we can obtain the first variation of the surface energy functional \eqref{eqn:strongenergy}:
\begin{equation}\label{eqn:strongvariaton}
\delta W^{\varepsilon}_{\rm {reg}}(\Gamma;\vec V) = \int_{\Gamma}\Bigl[-\partial_s\boldsymbol{\xi}^{\perp}\cdot \vec n-\varepsilon^2\Bigl(\frac{\kappa^3}{2}+\partial_{ss}\kappa\Bigr)\Bigr]\,(\vec V_0 \cdot \vec n)\;ds+\Bigl[(\xi_2-(\gamma_{_{\subVS}}-\gamma_{_{\subFS}})+\varepsilon^2\,\partial_s\kappa\,n_1)(\vec V_0\cdot\vec e_1)\Bigr]\Big|_{s=0}^{s=L}.
\end{equation}
Thus, the first variation of the strongly anisotropic surface energy functional with respect to the curve $\Gamma$ and the two contact points can be written as
\begin{eqnarray}
&&\frac{\delta W^{\varepsilon}_{\rm {reg}}}{\delta\Gamma}=-(\partial_s\boldsymbol{\xi})^{\perp}\cdot\vec n-\varepsilon^2(\frac{\kappa^3}{2}+\partial_{ss}\kappa),\\
&&\frac{\delta W^{\varepsilon}_{\rm {reg}}}{\delta x_c^l}=-\Bigl(\xi_2-(\gamma_{_{\subVS}}-\gamma_{_{\subFS}})+\varepsilon^2\,\partial_s\kappa\,n_1\Bigr)\Big|_{s=0},\\
&&\frac{\delta W^{\varepsilon}_{\rm {reg}}}{\delta x_c^r}=\Bigl(\xi_2-(\gamma_{_{\subVS}}-\gamma_{_{\subFS}})+\varepsilon^2\,\partial_s\kappa\,n_1\Bigr)\Big|_{s=L}.
\end{eqnarray}
Similar to the weakly anisotropic case and by using the same dimensionless scale units, we can obtain a dimensionless sharp-interface model again~\cite{Jiang16,Bao17} for solid-state dewetting of thin films with strongly anisotropic surface energy via a $\boldsymbol{\xi}$-vector formulation, which can be written as follows (for simplicity, we still use the same notations for the variables):
\begin{eqnarray}\label{eqn:strong1}
&&\partial_t\vec X=\partial_{ss}\mu\;\vec n, \qquad 0<s<L(t), \qquad t>0,\\
\label{eqn:strong2}
&&\mu = -\left[\partial_s\boldsymbol{\xi}\right]^{\perp}\cdot\vec n-\varepsilon^{2}\Big(\frac{\kappa^{3}}{2}+\partial_{ss}\kappa\Big),\quad
\kappa=-\left(\partial_{ss}\vec{X}\right)\cdot\vec{n},\quad \boldsymbol{\xi} = \nabla\hat{\gamma}(\vec p)\Big|_{\vec p=\vec n};
\end{eqnarray}
where $\Gamma:=\Gamma(t)=\vec X(s,t)=(x(s,t),y(s,t))$ represents the moving film/vapor interface, $s$ is the arc length or distance along the interface, $t$ is the time, $\vec n=(n_1,n_2)=(-\partial_{s}y, \partial_{s}x)$ is the interface outer unit normal vector, $\mu:=\mu(s,t)$ is the chemical potential, $\boldsymbol{\xi}=(\xi_1,\xi_2)$ is the Cahn-Hoffman vector and  $L:=L(t)$ represents the total length of the moving interface, $\varepsilon$ is a small regularization parameter. The initial condition is given as
\begin{equation}\label{init}
\vec{X}(s,0):=\vec{X}_0(s)=(x(s,0),y(s,0))=(x_0(s),y_0(s)), \qquad 0\le s\le L_0:=L(0),
\end{equation}
satisfying $y_0(0)=y_0(L_0)=0$ and $x_0(0)<x_0(L_0)$, and the boundary conditions are:
\begin{itemize}
\item[(i)] contact point condition:
\begin{equation}
y(0,t)=0, \qquad y(L,t)=0,\qquad t\geq 0,
\label{eqn:sBC1}
\end{equation}
\item[(ii)] relaxed contact angle condition:
\begin{equation}\label{eqn:sBC2}
\frac{d x_c^l}{d t}=\eta\Big(\xi_2-\sigma+\varepsilon^{2}\partial_{s}\kappa\,n_1\Big)\Big|_{s=0},\qquad
\frac{d x_c^r}{d t}=- \eta\Big(\xi_2-\sigma+\varepsilon^{2}\partial_{s}\kappa\,n_1\Big)\Big|_{s=L},
\end{equation}
\item[(iii)] zero-mass flux condition:
\begin{equation}
\partial_s \mu(0,t)=0, \qquad \partial_s \mu(L,t)=0,\qquad t\geq 0,
\label{eqn:sBC3}
\end{equation}
\item[(iv)] zero-curvature condition:
\begin{equation}
\kappa(0,t)=0,\qquad\kappa(L,t)=0,\qquad t\geq 0.
\label{eqn:sBC4}
\end{equation}
\end{itemize}

Following with the similar process as the weakly anisotropic case, the total area/mass conservation and energy dissipation properties under the strongly anisotropic case can be also rigorously proved. Here we omit the details.

\subsection{A discretization by PFEM and numerical simulations}

In order to obtain the variational formulation for equations \eqref{eqn:strong1}-\eqref{eqn:strong2} with the boundary conditions \eqref{eqn:sBC1}-\eqref{eqn:sBC4}, we need to introduce a new vector field $\vec I$. Here we assume that $\vec I$ is parallel to $\vec n$, thus we are able to rewrite the equations as the following form
\begin{eqnarray}
&&\partial_{t}\vec X\cdot\vec n=\partial_{ss}\mu,\qquad
\mu\;\vec n =-\left(\partial_{s}\boldsymbol{\xi}\right)^{\perp}-\varepsilon^2\vec I,\\
&&\vec I\cdot\vec n=\Bigl(\frac{1}{2}\kappa^3+\partial_{ss}\kappa\Bigr),\qquad
\kappa\;\vec n=-\partial_{ss}\vec X.
\end{eqnarray}
By using the integration by parts, the weak formulation for solid-state dewetting problems with strongly
anisotropic surface energy can be stated as the following variational problem: given the initial curve $\Gamma(0)=\vec X(\rho,0), \rho \in I$, for every time $t \in (0,T]$, find the evolution curves $\Gamma(t)=\vec X(\rho,t)\in H_{a,b}^1(I)\times H_{0}^{1}(I)$ with the $x$-coordinate positions of moving contact points $a=x_c^l(t)\leq x_c^r(t)=b$, the chemical potential $\mu(\rho,t)\in H^1(I)$, $\vec I(\rho,t)\in H^1(I)\times H^1(I)$ and the curvature $\kappa(\rho,t)\in H^1_0(I)$ such that
\begin{subequations}
\begin{align}
&\big<\partial_{t}\vec{X},~\varphi\vec{n}\big>_{\Gamma}+\big<\partial_{s}\mu,~\partial_{s}\varphi\big>
_{\Gamma}=0,\quad \forall\; \varphi\in H^{1}(I), \label{eqn:strongvf1}\\
&\big<\mu\vec n,~\boldsymbol{\omega}_1\big>_{\Gamma}-\big<\boldsymbol{\xi}^{\perp},~\partial_s\boldsymbol{\omega}_1\big>_{\Gamma}+\varepsilon^2\big<\vec I,~\boldsymbol{\omega}_1\big>_{\Gamma} +(\boldsymbol{\xi}^{\perp}\cdot\boldsymbol{\omega}_1)|_{s=0}^{s=L(t)}=0,\qquad\forall\,\boldsymbol{\omega}_1\in (H^1(I))^2,\label{eqn:strongvf2}\\
&\big<\vec I,~\vec n\phi\big>_{\Gamma}-\frac{1}{2}\big<\kappa^3,~\phi\big>_{\Gamma}+\big<\partial_s\kappa,~\partial_s\phi\big>_{\Gamma}=0,\quad\forall\,\phi\in H^1_0(I),\label{eqn:strongvf3}\\
&\big<\kappa\vec n,~\boldsymbol{\omega_2}\big>_{\Gamma}-\big<\partial_s\vec X,~\partial_s\boldsymbol{\omega_2}\big>_{\Gamma} =0,\quad\forall\,\boldsymbol{\omega_2}\in (H^1_0(I))^2.\label{eqn:strongvf4}
\end{align}
\end{subequations}

Following with the designing idea for weakly anisotropic case, we can propose a discrete semi-implicit PFEM for solving the above sharp-interface model with strongly anisotropic surface energies, according to the above variational formulation. We omit the details because the scheme is directly an easy extension of the PFEM for weakly anisotropic case.
\begin{figure}[!htp]
\centering
\includegraphics[width=0.7\textwidth]{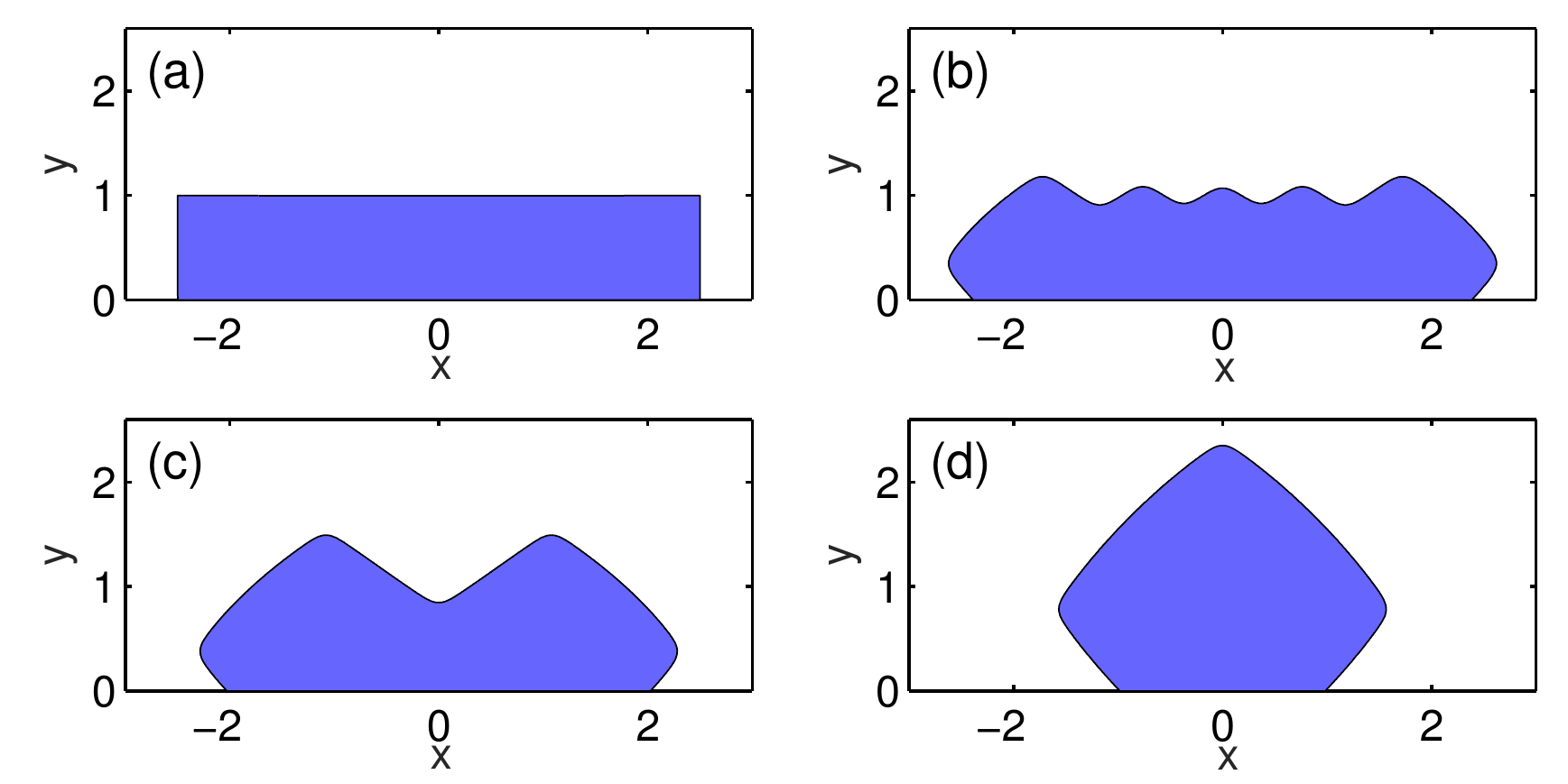}
\caption{\label{fig:stevolution} Several snapshots in the geometric evolution of a small initially rectangular island to the equilibrium shape (blue line). (a) $t=0$, (b) $t=0.02$, (c) $t=0.04$, (d) $t=10.32$, where $\gamma(\theta)=1+0.2\cos(4\theta)$, $\sigma=\cos(2\pi/3),\varepsilon=0.1$.}
\end{figure}

The first numerical example for strongly anisotropic case is set up as follows. The initial thin film is chosen as a rectangular island with width $L=5$ and height $h=1$. The anisotropic surface energy is given by the $4$-fold crystalline surface energy defined in Eq.~\eqref{eqn:kfoldenergy} with $\beta=0.2$. The computational parameters are chosen as $\sigma=\cos(2\pi/3),\varepsilon=0.1$. As depicted in Fig.~\ref{fig:stevolution}, it shows the kinetic evolution process of an initial rectangle island film towards its equilibrium shape. We can observe that wavy structure first appears during the evolution (Fig.~\ref{fig:stevolution}(b)), and eventually the island film evolves into a almost ``faceting'' shape with three regularized rounded corners (Fig.~\ref{fig:stevolution}(d)).

\begin{figure}[!htp]
\centering
\includegraphics[width=0.7\textwidth]{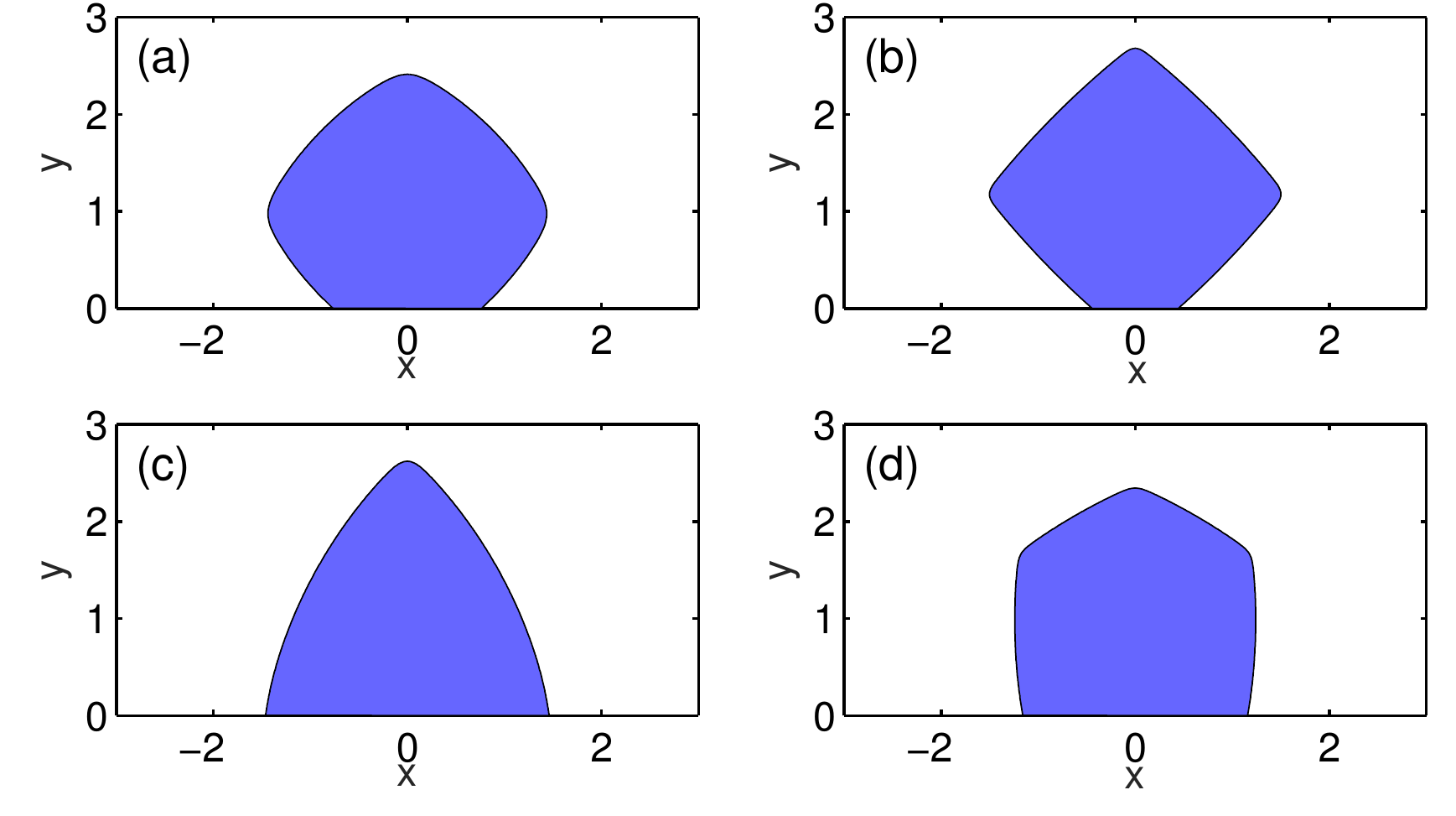}
\caption{\label{fig:stmb} Equilibrium shapes for a small, initially rectangular island under several different crystalline fold of symmetry $k$ and degree of anisotropy $\beta$ respectively, where the surface energy density are chosen as the smooth $k$-fold anisotropy: (a) $k=4,\beta=0.1$; (b) $k=4,\beta=0.3$; (c) $k=3,\beta=0.3$; (d) $k=6,\beta=0.1$, and $\sigma=\cos(3\pi/4),\varepsilon=0.1$.}
\end{figure}

As illustrated in Fig.~\ref{fig:stmb}, we also compute the equilibrium shape for a small, initially rectangular island film with different strongly anisotropic smooth $k$-fold surface energy under the parameters $\sigma=\cos(3\pi/4),\varepsilon=0.1$. Fig.~\ref{fig:stmb}(a)-(b) depict the effect of increasing the degree of anisotropy from $\beta=0.1$ to $\beta=0.3$ for $4$-fold crystalline symmetry, and we can see that the equilibrium shape becomes more and more faceting. The effect of increasing the rotational symmetry, which is reflected from increasing the parameter $k$, can be observed from Fig.~\ref{fig:stmb}(c) ($k=3,\beta=0.3$) to Fig.~\ref{fig:stmb}(d) ($k=6,\beta=0.1$).

\section{Summary and Conclusions}

Based on a Cahn-Hoffman $\boldsymbol{\xi}$-vector formulation, we have proposed sharp-interface models and corresponding parametric finite element numerical schemes for simulating solid-state dewetting of thin films with anisotropic
surface energies in two dimensions, and these sharp-interface approaches (including mathematical models and numerical schemes) can handle with any type of surface energy density no matter what its form is ($\gamma(\vec n)$ or $\gamma(\theta)$).

First, by using the thermodynamic variation, we rigorously derive the first variation of the total interfacial energy functional of solid-state dewetting system via the smooth vector-field perturbation method, and obtain a sharp-interface model for describing the kinetic evolution of solid-state dewetting of thin films with weakly anisotropic surface energies. The governing equation is described by surface diffusion and contact line migration. Second, a variational formulation via the $\boldsymbol{\xi}$-vector is proposed for the sharp-interface governing equation, and we prove that its fully-discrete scheme (i.e., PFEM) is well-posed. Then, by performing numerical simulations for different types of surface energy anisotropy, we examine several specific evolution processes for solid-state dewetting island thin films, e.g., equilibrium shapes of small islands, pinch-off of large islands and power-law retraction dynamics of semi-infinite island films. Finally, we also include the strong surface energy anisotropy effect into the sharp-interface model and its corresponding numerical scheme via the $\boldsymbol{\xi}$-vector formulation.

It should be noted that, compared to the previous derivations of the sharp-interface models by using scalar-field perturbation method~\cite{Wang15,Jiang16,Bao17b}, the vector-field perturbation method via a Cahn-Hoffman $\boldsymbol{\xi}$-vector formulation is much simpler and its intrinsic physical meaning of the variational structure is also much more direct and clear. Furthermore, based on this $\boldsymbol{\xi}$-vector formulation, the parametric finite element numerical scheme for solving the sharp-interface models shares more advantages than the previous one proposed by us~\cite{Bao17}. While the present sharp-interface approaches are for two dimensions, our approaches (including models and numerical schemes) via the $\boldsymbol{\xi}$-vector formulation can be directly generalized to three dimensions~\cite{Bao18,Zhao17thesis}. We believe that our approach can shed some light on understanding this kind of thermodynamic variation problem (i.e., the energy functional depends on an open curve or open surface), and offer a convenient tool to designing the mathematical model and its corresponding numerical scheme for this kind of problems.

\section*{Acknowledgements}

The authors would like to thank Prof.~Weizhu Bao and Prof.~David J. Srolovitz for fruitful discussions. We acknowledge the supports from the National Natural Science Foundation of China 11401446 and 11571354 (W.J.) and the Ministry of Education of Singapore grant R-146-000-247-114 (Q.Z.). The first author also thanks the support from the National Natural Science Foundation of China 91630207 when he was visiting Beijing Computational Science Research Center in 2018.

\section*{References}
\bibliographystyle{elsarticle-num}
\bibliography{thebib}
\end{document}